\documentclass{article}
\usepackage{graphicx}

\usepackage[utf8]{inputenc} 
\usepackage{icomma,amsmath,hyperref,graphicx} 
\usepackage[final]{pdfpages} 
\usepackage[left=3.2cm,right=3.2cm,bottom=1.2in, top=1.2in]{geometry} 

\usepackage{amsfonts}

\usepackage{makecell}
\usepackage{lipsum}
\usepackage{amssymb}
\usepackage{multicol, caption}

\usepackage[utf8]{inputenc}

\usepackage{mathtools}

\usepackage[export]{adjustbox}
\usepackage{wrapfig}

\usepackage{mwe}
\usepackage{braket}

\usepackage{isotope}
\usepackage{chemformula}

\usepackage{cite}

\let\OLDthebibliography\thebibliography
\renewcommand\thebibliography[1]{
  \OLDthebibliography{#1}
  \setlength{\parskip}{1pt}
  \setlength{\itemsep}{0pt plus 0.3ex}
}  

\usepackage{setspace}
\makeatletter
\newcommand*{\rom}[1]{\expandafter\@slowromancap\romannumeral #1@} 

\date{}
\begin{document}

\title{Understanding and Optimizing the Sensitization of Anatase Titanium Dioxide Surface with Hematite Clusters}

\author{Kati Asikainen\ $^{1*}$, Matti Alatalo $^1$, Marko Huttula $^1$, \\
B. Barbiellini $^2$ and S. Assa Aravindh\ $^{3*}$ \\[0.3cm] \normalsize ${}^1$\textit{Nano and Molecular Systems Research Unit, University of Oulu, FI-90014, Finland} \\[0.2cm] \normalsize ${}^2$\textit{Lappeenranta-Lahti University of Technology (LUT), FI-53851 Lappeenranta, Finland}\\
\normalsize ${}^3$\textit{Sustainable Chemistry and MME, Faculty of Technology, University of Oulu, FI-90014, Finland}\\
\begin{tabular}{l}
\\[0.1cm]
\small ${}^*$Corresponding author: Kati.Asikainen@oulu.fi \\
\small ${}^*$Corresponding author: Assa.SasikalaDevi@oulu.fi
\end{tabular}
}

\maketitle

\begin{abstract}
The presence of  hematite (\ch{Fe2O3}) clusters at low coverage on titanium dioxide (\ch{TiO2}) surface has been observed to enhance photocatalytic activity, while excess loading of hematite is detrimental. We conduct a comprehensive density functional theory study of \ch{Fe2O3} clusters adsorbed on the anatase \ch{TiO2} (101) surface to investigate the effect of \ch{Fe2O3} on \ch{TiO2}. Our study shows that \ch{TiO2} exhibits improved photocatalytic properties with hematite clusters at low coverage, as evidenced by a systematic study conducted by increasing the number of cluster adsorbates. The adsorption of the clusters generates impurity states in the band gap improving light absorption and consequently affecting the charge transfer dynamics. Furthermore, the presence of hematite clusters enhances the activity of \ch{TiO2} in the hydrogen evolution reaction. The Fe valence mixing present in some clusters leads to a significant increase in \ch{H2} evolution rate compared with the fixed +3 valence of Fe in hematite. We also investigate the effect of oxygen defects and find extensive modifications in the electronic properties and local magnetism of the \ch{TiO2} -\ch{Fe2O3} system, demonstrating the wide-ranging effect of oxygen defects in the combined system.
\end{abstract}


\vspace{0.3cm}

\section{Introduction}
Photocatalysis emerges as a promising technique to address contemporary energy challenges. In particular, photocatalytic water splitting represents a highly sustainable approach to produce green hydrogen and oxygen by dissociating water using natural light. This environmentally friendly method enables the conversion of solar energy into chemical energy without contributing to pollution. Semiconductors with band gaps ranging from 2.5 to 3.4 eV are favorable photocatalysts covering visible to ultraviolet (UV) spectra \cite{1}. Consequently, incident photons with energies greater than or equal to the band gap of the material liberate electrons to the conduction band (CB) and holes to the valence band (VB). These photogenerated charge carriers can migrate to the surface of the material participating in oxidation and reduction reactions. Amongst the semiconductors suitable for these purposes, \ch{TiO2} has been considered as a premier material since its discovery in 1972 by Fujishima and Honda \cite{2}. It possesses exceptional properties that make it attractive for photocatalysis. These properties include oxidation properties, physical and chemical stability, non-toxicity and abundance \cite{3,4,5,G}. Moreover, \ch{TiO2} possesses a band structure suitable for efficient water splitting, as it straddles the reduction (-4.44 eV) and oxidation potentials (-5.67 eV) of water \cite{G}. However, despite the favorable attributes, a large band gap, quick recombination of the electron-hole pairs and being inactive for overall water splitting in the absence of sacrificial reagents leads to a performance degradation of \ch{TiO2} \cite{5,6,R}. These factors limit the widespread utilization of \ch{TiO2} in photocatalytic applications. Therefore, attempts are made to overcome the shortcomings and enhance the performance of \ch{TiO2}. 

\ch{TiO2} exhibits three phases at atmospheric pressure: stable rutile, metastable anatase and brookite. The crystalline structure can strongly affect the photocatalytic activity of a material. Since the synthesis of pure brookite is difficult \cite{brookite}, extant investigations have mainly focused on rutile and anatase phases. Even though anatase possesses a larger band gap than rutile, anatase has been considered to have a superior photocatalytic performance \cite{A1, A2, A3, A4, A5}. The better performance is generally attributed to indirect band gap and lower effective mass of photogenerated electrons and holes, increasing the lifetime of electron-hole pairs and lowering their recombination rate \cite{A6}. However, researchers have also reported a good performance of rutile in special experimental synthesis conditions \cite{R1, R2}. Moreover, mixing anatase and rutile phase structures have  shown even better activity than pure anatase, attributed to interfacial charge transfer occurring between the two phases \cite{R2, M1}.

A variety of strategies have been proposed to modify both chemical and physical properties of pure \ch{TiO2}, and the construction of heterostructure systems has shown improvements in photocatalytic performance \cite{7}. It is recognized that by combining appropriate photocatalytic materials, heterojunction structures with matching band alignments can be formed. One promising candidate for combination with \ch{TiO2} is hematite ($\alpha$-\ch{Fe2O3}, hereafter referred to as \ch{Fe2O3}). It is the most stable phase of iron (\rom{3}) oxide, possessing a narrow band gap of 2.0-2.2 eV \cite{11,12}. Several studies on \ch{TiO2}-\ch{Fe2O3} interface have been carried out earlier. For instance Mei et al. \cite{Mei} and Singh et al. \cite{Singh} have synthesized \ch{TiO2}-\ch{Fe2O3} heterostructure using deposition methods, where the \ch{TiO2}(\ch{Fe2O3}) surface was fully covered and \ch{Fe2O3} and \ch{TiO2} are in direct contact. The results of Mei et al. suggested that \ch{Fe2O3} concentration plays a key role in light adsorption, and the photoresponse could be engineered by the amount of surface \ch{Fe2O3} in \ch{TiO2}.  However, many studies have found that the increased amount of \ch{Fe2O3} content is not leading to a superior photocatalytic activity. For instance, in the study of Mei \textit{et al.}, using the photoluminescence spectroscopy (PL), the PL intensity of TH0.5 sample was found to be the lowest, indicating the most efficient charge separation compared to the samples with lower or higher \ch{Fe2O3} content. Cao et al. \cite{14} have the same conclusion on \ch{Fe2O3} coated \ch{TiO2} using atomic layer deposition (ALD)  methods. They reported that 400 cycles of \ch{Fe2O3} coating led to better photocatalytic activity than 200, 600 and/or 800 cycles. In general, at high \ch{Fe2O3} content iron has been proposed to become a recombination center for the charge carriers, having a negative impact on catalytic activity \cite{8,15}. Singh et al. \cite{Singh} studied \ch{TiO2}-\ch{Fe2O3} heterostructure over pure \ch{Fe2O3} surface and a low performance in photoelectrochemical cells (PEC) was attributed to inefficient charge separation in the heterostructure. It is also worth noticing that in their computational investigation, they highlighted that anatase \ch{TiO2} (101) surface placed on top of the \ch{Fe2O3} surface was reconstructed during the geometry optimization. \ch{TiO2} was reported to exhibit an "amorphous" structure that resembled none of the stable crystal structures of \ch{TiO2}. The notable reconstruction may be attributed to the large lattice mismatch of the two surfaces. Further, investigations focusing on heterostructures with various \ch{Fe2O3} coverage and content on \ch{TiO2}, have shown that finding an optimal \ch{Fe2O3} content is essential in order to maximize the photocatalytic activity \cite{8,14,15,plasma,bacteria,Mei, Abbas}.  
For instance, Sun \textit{et al.} \cite{15} conducted an experimental study on anatase \ch{TiO2} surface modified with the hematite cluster for phenol degradation. They reported that small clusters enhanced the photocatalytic activity of \ch{TiO2} by improving its response to visible light, and charge carrier transfer and separation. In general, the observed enhancement in catalytic activity has been attributed to the formation of a heterojunction between \ch{TiO2} and \ch{Fe2O3} \cite{8,14,15,17,plasma,13,9}. Based on a thorough literature search, we can conclude that increased coverage of \ch{Fe2O3} is detrimental to photocatalytic performance of the material.

Defect engineering emerges as another effective method to modify the catalytic properties and mechanisms. In metal oxides, including \ch{TiO2}, oxygen vacancies are common defects \cite{19}. Resulting from an oxygen vacancy, impurity states are generally observed in the band gap of \ch{TiO2}, often manifested as Ti$^{3+}$ species. These impurity states contribute to a shift in the absorption spectrum and increased conductivity of \ch{TiO2} \cite{23,25}. Consequently, these can lead to enhanced photocatalytic activity, demonstrated by Sheiber \textit{et al.} \cite{26} for example. They reported that oxygen vacancies improved the photocatalytic activity of anatase \ch{TiO2} in water adsorption. Several experiments have also demonstrated an existence of oxygen defect states in \ch{Fe2O3}-\ch{TiO2} and Fe-\ch{TiO2} systems, and the states have been attributed to oxygen vacancies in the \ch{TiO2} lattice \cite{C1, C2, C3, C4}. Spectroscopy techniques, such as X-ray photoelectron spectroscopy (XPS), have been used to confirm oxygen vacancy defects in the systems. The studies have reported a successful incorporation of Fe$^{3+}$ into the \ch{TiO2} lattice, and due to the charge compensation mechanism, Fe$^{3+}$ species results in the formation of oxygen vacancies. For instance, Zhu \textit{et al.} \cite{C4} proposed that the oxygen vacancies could contribute to a reduction in recombination rate. Correspondingly, Bootluck \textit{et al.} \cite{plasma} suggested that oxygen vacancies contribute to enhanced photocatalytic activity. They reported that resulting from the Ar-plasma treatment, oxygen vacancies were produced in the \ch{Fe2O3}-\ch{TiO2} nanocomposites, contributing to  improved photocatalytic activity compared to untreated \ch{Fe2O3}-\ch{TiO2}. It is also demonstrated that \ch{Fe2O3}-\ch{TiO2} may be suitable material for hydrogen evolution reaction (HER). The improved photocatalytic activity of \ch{Fe2O3}-\ch{TiO2} nanocomposites was attributed to a lower charge transfer resistance, which has led to better charge separation and reduced recombination rate of electron-hole pairs compared to bare \ch{TiO2}. It is worth noting that the study did not provide direct comparison between the HER activity of \ch{Fe2O3}-\ch{TiO2} and bare \ch{TiO2}.

  
Inspired by the investigations  of Sun \textit{et al.} \cite{8},  we employ first principles calculations to investigate the adsorption of hematite clusters on anatase \ch{TiO2} (101) surfaces and its effect on photocatalytic properties. The choice of the anatase phase was motivated by its superior photocatalytic activity compared to other polymorphs, as discussed earlier. Moreover, studies have reported that the (101) plane of anatase possesses the highest photocatalytic activity than other crystal orientations \cite{31, 30}. In order to improve the photocatalytic properties of \ch{TiO2} Sun \textit{et al.} \cite{15} highlighted the advantage of small \ch{Fe2O3} cluster size and low \ch{Fe2O3} coverage on anatase \ch{TiO2}. Building upon this, we introduce small hematite clusters, (\ch{Fe2O3})$_n$ with $n=1,2,3$ at the \ch{TiO2} surface. We also delve into defect engineering as another effective method to modify the properties of the heterostructures. More precisely, we introduce an oxygen defect in the heterostructure of \ch{TiO2} and (\ch{Fe2O3})$_1$. Finally, to assess the photocatalytic performance, we focus on evaluating the suitability of the investigated heterostructures for HER. Our aim is to understand the effects brought about by the hematite clusters and further oxygen defect towards the electronic properties of \ch{TiO2}, due to the change in carrier concentration compared to the pristine surface. We also demonstrate that coating the anatase titanium dioxide with mixed-valence iron containing O-ligands using various hematite cluster coverings represents an effective approach to enhance photocatalytic hydrogen production.
 

\section{Computational methodology}
First principles calculations based on density functional theory (DFT) were performed using Vienna Ab initio Simulation Package (VASP) \cite{34,35,36} including the spin polarization. The exchange-correlation potential was described by generalized gradient approximation (GGA) parameterized by Perdew-Burke-Ernzerhof (PBE) functional \cite{37}. The Hubbard correction was employed according to Dudarev \textit{et al.} \cite{38} to describe localized $d$ electrons of titanium and iron in order to obtain more realistic electronic and magnetic properties. Within the GGA+U method we adopted the corrections of U$_\mathrm{eff} = 4.5$ eV (U$=4.5$ eV and J$=0$ eV) \cite{39} for the Ti (table S1) and U$_\mathrm{eff} = 4.0$ eV (U$=4.0$ eV and J$=0$ eV) \cite{40} for the Fe. In some studies the Hubbard correction has also been applied to O 2p states to further improve the description of Ti-O bonds and obtain a larger opening of the band gap for \ch{TiO2} \cite{UO}. Based on our results, in  bulk \ch{TiO2}, applying the U correction to the O 2p states did not provide a significant improvement in the band gap (table S1), and therefore we chose to consider the Hubbard correction only for Ti 3d states. The 1st Brillouin zone was sampled according to the Monkhorst-Pack scheme \cite{42}. The Projector augmented wave (PAW) method \cite{41} was employed to describe the electron-ion potential with the plane wave energy cutoff of 650 eV (figure S1). The atomic relaxations were performed until the forces and energies were less than 0.001 eV/{\AA} and $10^{-6}$\ eV, respectively. We used VESTA \cite{43} for visualization and VASPKIT \cite{vaspkit} for post-processing of the data from the VASP calculations. 

To simulate the pristine anatase \ch{TiO2} (101) surface we constructed a slab of 4 layers containing 192 atoms (64 Ti atoms and 128 O atoms) with lattice parameters of 10.32\ \AA\, 15.25\ \AA\ and 35.68\ \AA\ in $a$, $b$ and $c$ directions (figure 2). To prevent interaction between periodic images, a vacuum space of 20\ \AA\ was applied along the $c$-axis. The heterostructures were formed by placing the hematite clusters (\ch{Fe2O3})$_{1,2,3}$ on the \ch{TiO2} surface. The models were denoted as (\ch{Fe2O3})$_1$/\ch{TiO2}, (\ch{Fe2O3})$_2$/\ch{TiO2} and (\ch{Fe2O3})$_3$/\ch{TiO2}, and only these three heterostructures were considered in this study. K-point sampling of $3\times 2\times 1$ and Gaussian smearing of 0.05 eV were used for the \ch{TiO2} surface and heterostructures. To investigate the effect of an oxygen defect, several defect sites, shown in figure \ref{fig:sites-O-vacs}, were created in the (\ch{Fe2O3})$_1$/\ch{TiO2}. An oxygen atom was removed either from the \ch{TiO2} (surface or subsurface vacancy) or from the (\ch{Fe2O3})$_1$ cluster. We created an oxygen vacancy in the surface layer of \ch{TiO2} by removing a twofold-coordinated O atom, not forming a bond with the cluster. Three subsurface oxygen atoms were removed at different distances from the cluster in the first subsurface layer. Lastly, two different oxygen defects were created in  (\ch{Fe2O3})$_1$ cluster. The surface vacancy was labeled as O$_\mathrm{v}$, the subsurface vacancies as O$_\mathrm{sv1}$, O$_\mathrm{sv2}$ and O$_\mathrm{sv3}$, and the oxygen defects located in the (\ch{Fe2O3})$_1$ cluster as O$_\mathrm{c1}$ and O$_\mathrm{c2}$. The defective heterostructures were further denoted as (\ch{Fe2O3})$_1$/\ch{TiO2}-O$_\mathrm{vac}$ where O$_\mathrm{vac}$ specifies the oxygen defect. In addition, to comprehensively investigate the effect of hematite clusters on the HER activity, we also considered the \ch{Fe2O3} surface in our calculations (figure S2). Thus, the HER activity of heterostructures was compared with both pristine \ch{TiO2} and \ch{Fe2O3} surfaces. For hydrogen adsorption a supercell of $(2\times 2 \times 1)$ of \ch{Fe2O3}(0001) surface with a single Fe-termination with a vacuum thickness of around 20 \AA\ was taken, and k-point sampling of $3 \times 3 \times 1$ was used in the calculations.

The adsorption energy of the (\ch{Fe2O3})$_n$ cluster at the \ch{TiO2} surface was calculated from the formula
\begin{align}
E^{\mathrm{Ads}}=E((\ch{Fe2O3})_n/\ch{TiO2})-E(\ch{TiO2})-E((\ch{Fe2O3})_n),
\end{align}
where $E((\ch{Fe2O3})_n/\ch{TiO2})$ is the total energy of the  heterostructures and $E(\ch{TiO2})$ and $E((\ch{Fe2O3})_n)$ are the total energies of the pristine \ch{TiO2} surface and (\ch{Fe2O3})$_n$ cluster, respectively. To investigate the stability of oxygen defects, we calculated the formation energy of the defects using the equation
\begin{align}
E^{\mathrm{Form}}=E((\ch{Fe2O3})_1/\ch{TiO2}-\mathrm{O_{vac}})-E((\ch{Fe2O3})_1/\ch{TiO2})-\frac{1}{2} E(\ch{O2}),
\end{align}
where the first term is the total energy of the defective heterostructure, in which the O$_\mathrm{vac}$ specifies the investigated oxygen defect. $\frac{1}{2}E(\ch{O2})$ is the chemical potential of an oxygen atom that is half of the total energy of an isolated oxygen molecule \ch{O2}. To evaluate the HER activity, we attached a hydrogen atom on the pristine \ch{TiO2} and \ch{Fe2O3} surfaces, and  also on defect-free and defective heterostructures (figure S3 and S4), and calculated the adsorption energy of hydrogen. In general, the free energy of hydrogen adsorption is accepted to be a descriptor for hydrogen-evolving catalysts. We also calculated the work function for the heterostructures. Work function is essentially the energy needed to introduce carriers to the surface and will be affected by doping and the presence of adsorbates. It is an essential parameter in understanding the interaction between the hematite clusters and \ch{TiO2} and the effect of oxygen defect on the surface properties, and it is calculated by subtracting the Fermi energy $E_F$ from the vacuum energy $E_V$,
\begin{align}
\Phi=E_V-E_F.
\end{align} 

The electronic structure of the systems investigated was examined through the density of states (DOS). To analyze the charge distribution and charge transfer quantitatively, we performed the Bader analysis \cite{45,46,47,48}. A negative Bader charge on an atom refers to electron gain and positive value to electron loss. The charge density differences are also calculated as 
\begin{align}
\Delta \rho(\textbf{r})=\rho_{\mathrm{AB}}(\textbf{r})-\rho_{\mathrm{A}}(\textbf{r})-\rho_{\mathrm{B}}(\textbf{r})
\end{align}
where $\rho_{\mathrm{AB}}(\textbf{r})$ is the total charge density of the heterostructure, and $\rho_{\mathrm{A}}(\textbf{r})$ and $\rho_{\mathrm{B}}(\textbf{r})$ are the total charge densities of the \ch{TiO2} surface and the (\ch{Fe2O3})$_n$ cluster with atoms in exactly the same sites as they occupy in the heterostructure. Since \ch{Fe2O3} is a magnetic material, we also calculated the spin density difference. It is calculated from the same equation as charge density difference by replacing charge density $\rho(\textbf{r})$ by spin density $s(r)$. In the charge density difference (CDD) plots yellow refers to charge accumulation and cyan refers to charge depletion while in the spin density difference (SDD) plot the orange shows excess spin up polarization and turquoise shows excess spin down polarization. We set the isosurface value of 0.005\ \textit{e}\ \AA$^{-1}$ \ in all CDD and SDD plots.  

\section{Results}
\vspace{-0.1cm}
\subsection{Bulk parameters}
Initially, we calculated the bulk parameters of \ch{Fe2O3} and \ch{TiO2} (figure S5). The lattice constants were determined to be $a=b=4.780$\ \AA \ and $c =13.323$\ \AA \ and $a=b=3.98$\ \AA \ and $c=9.56$\ \AA \ for \ch{Fe2O3} and \ch{TiO2}, respectively. These results are consistent with experimental measurements \cite{Fe2O3-parameters, 44}. The DOS of the bulk \ch{Fe2O3} and bulk \ch{TiO2} are in figure S6, showing the electronic structure of the materials. Using the GGA functional \ch{Fe2O3} was predicted to exhibit metallic behaviour whereas \ch{TiO2} possessed a band gap of 1.7\ eV. The results demonstrate that the standard GGA functional is incapable of accurately describing both \ch{Fe2O3} and \ch{TiO2} because of the localized d electrons. This is a well-known issue in predicting the electronic structure of transition metals oxides \cite{I1, I2, I3, I4, I5}. Previously, for instance, Labat \textit{et al.} \cite{Labat}, Mattioli \textit{et al.} \cite{I2} and Di Valentin \textit{et al.} \cite{DiValentin} have reported the band gaps of 2.08 eV, 2.16 eV and 2.19 eV for \ch{TiO2} while Cococcioni \textit{et al.} \cite{I3} found a zero band gap for \ch{Fe2O3} with the GGA. Upon employing the Hubbard correction we obtained a band gap of 1.16\ eV for \ch{Fe2O3} and 2.3\ eV for \ch{TiO2}, which are fairly consistent with the reported theoretical values \cite{I2, Fe2O3-DOS1, Fe2O3-DOS2, Fe2O3-DOS3, P1, HSE2}, showing improved description of the materials with the GGA+U. Previously, both materials have also been investigated using the hybrid functionals which, however, tend to overestimate the band gap value. Meng \textit{et al.} \cite{HSE1} calculated the band gap of 2.41 eV for \ch{Fe2O3} using the HSE functional with mixing parameter of $a=0.15$, and Yamamoto \textit{et al.} \cite{HSE2} and De$\Acute{\mathrm{a}}$k \textit{et al.} \cite{HSE3} have reported a band gap of 3.37 eV and 3.58 eV using the HSE06 functional. Even though hybrid functionals are generally more accurate for semiconductors, the GGA+U provides a reasonable compromise between the accuracy and computational cost in band gap calculations. Therefore, based on the results, we conclude that the selected methods are sufficient for describing the electronic properties of both materials.

\subsection{Freestanding clusters and \ch{TiO2} surface}
\begin{figure}[h!]\centering
\includegraphics[width=0.75\linewidth]{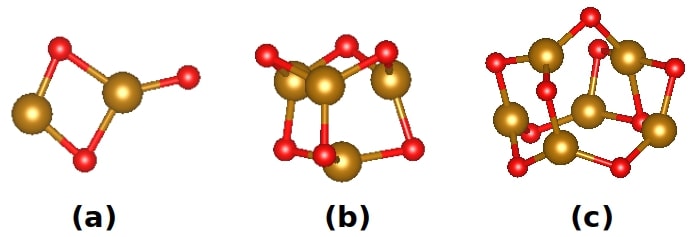}
\caption{The optimized structure of (a) (\ch{Fe2O3})$_1$, (b) (\ch{Fe2O3})$_2$ and (c) (\ch{Fe2O3})$_3$ clusters.}
\label{fig:clusters}
\end{figure}  
\begin{figure}[h!]\centering
\includegraphics[width=0.9\linewidth]{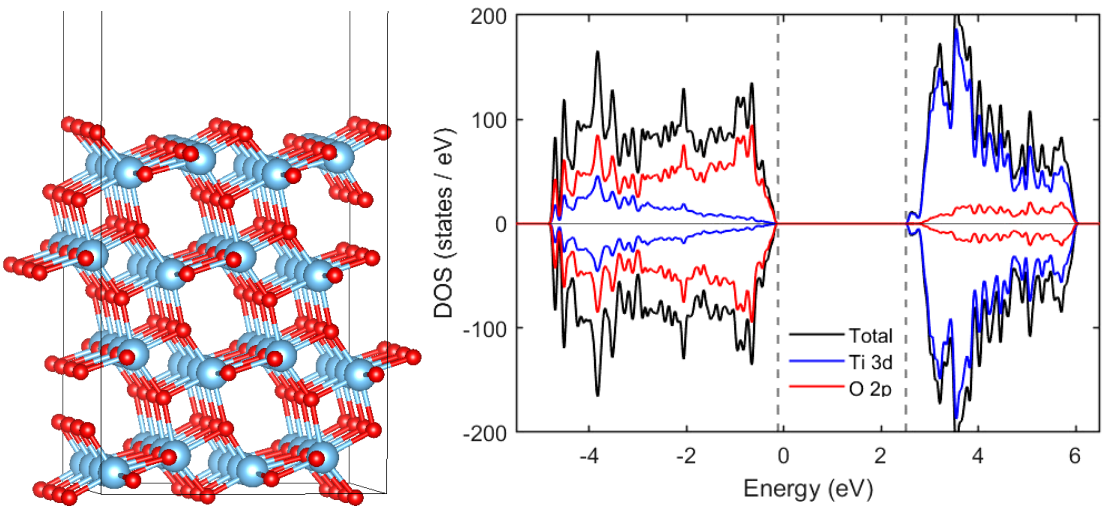}
\caption{The surface model and computed DOS of anatase \ch{TiO2}. A reasonable band gap of approximately 2.5 eV was found for the \ch{TiO2} surface. Fermi level is set to  zero energy.}
\label{fig:DOS-surface}
\end{figure}
Before investigating the heterostructures, we simulated the freestanding hematite clusters and pristine \ch{TiO2} surface. The relaxed structures of  (\ch{Fe2O3})$_1$, (\ch{Fe2O3})$_2$ and (\ch{Fe2O3})$_3$ are shown in figure \ref{fig:clusters}. The (\ch{Fe2O3})$_1$ showed a planar structure whereas the (\ch{Fe2O3})$_2$ and (\ch{Fe2O3})$_3$ had a cage-like structure. The Fe-O distances were ranging from 1.65 to 1.87 \AA\ in the (\ch{Fe2O3})$_1$, from 1.73 to 1.83 \AA\ in the (\ch{Fe2O3})$_2$ and from 1.69 to 1.99 \AA\ in the (\ch{Fe2O3})$_3$ (table S2). These geometries and Fe-O distances compared well with previous works \cite{Moniz, 51, Cluster1, Cluster2, TM-FexOy}, providing a good starting point for the rest of the calculations. Furthermore, we evaluated the oxidation state of iron in the free-standing clusters. Details of Bader charges and magnetic moments of atoms in the clusters are provided in table S3. According to the Bader analysis, Fe atoms exhibited a gain of charge which was depleted from O atoms. Ferromagnetic configuration was indicated for all the three clusters. The total magnetic moment of the (\ch{Fe2O3})$_1$ was 9.28$\ \mathrm{\mu_B}$, and spin magnetic moments of Fe atoms were 3.32$\ \mathrm{\mu_B}$ and 3.60$\ \mathrm{\mu_B}$, suggesting co-existence of Fe$^{2+}$ and Fe$^{3+}$ oxidation states \cite{TM-FexOy}. Bader charges of 1.08 \textit{e} and 1.46 \textit{e} supported the presence of mixed-valence Fe \cite{55}. In the (\ch{Fe2O3})$_2$ Bader charges of Fe atoms were around 1.30 \textit{e} for all four Fe atoms whereas magnetic moments of the Fe atoms were  3.70 $\ \mathrm{\mu_B}$, 3.32$\ \mathrm{\mu_B}$, 2.89$\ \mathrm{\mu_B}$ and 2.85$\ \mathrm{\mu_B}$. In the (\ch{Fe2O3})$_3$ the Bader charge and magnetic moment of Fe atoms varied in the range of 1.03 to 1.53 \textit{e}, and 2.97 to 4.04$\ \mathrm{\mu_B}$. The reduction in magnetic moment may imply lower oxidation state than Fe$^{3+}$ for Fe in these clusters.

We proceeded with the optimization of anatase \ch{TiO2} (101) surface. There are four types of atoms at the surface: two- and threefold coordinated oxygen atoms (O$_{2c}$ and O$_{3c})$ and five- and sixfold coordinated Ti atoms (Ti$_{5c}$ and Ti$_{6c}$). The calculated O-Ti bond lengths were 1.87\ \AA \ for O$_{2c}$-Ti$_{6c}$, 1.85\ \AA \ for O$_{2c}$-Ti$_{5c}$ and 2.00\ \AA \ for O$_{3c}$-Ti$_{6c}$. In addition, we found Ti$_{5c}$-Ti$_{5c}$ bonds to be 3.81\ \AA . The calculated lattice parameters were  $a=b=3.88$\ \AA \ and $c=9.52 $\ \AA. The electronic structure of the pristine \ch{TiO2} surface, calculated using the GGA+U, is shown in figure \ref{fig:DOS-surface}. The appearance of shoulder features in the CB edge is characteristic of the anatase phase of \ch{TiO2}. The valence band maximum (VBM) and conduction band minimum (VBM) were located at -0.05 eV and 2.44 eV, respectively, resulting in a band gap of approximately 2.5 eV. This agrees with the reported results \cite{Moniz, 52}. The band gap energy is in the visible light region indicating that the photoactivation of anatase surface could be achieved by visible light radiation with wavelength up to 500\ nm. Due to the dangling bonds the surface band gap of \ch{TiO2} is generally lower than the bulk band gap (3.0-3.3 eV). Moreover, the work function of the surface was calculated to be 7.23 eV. In general, \ch{TiO2} possesses a high work function of 5-6 eV \cite{54,WF}. Compared to this our calculations moderately overestimated the work function.

\begin{figure}[h!]\centering
\includegraphics[width=0.95\linewidth]{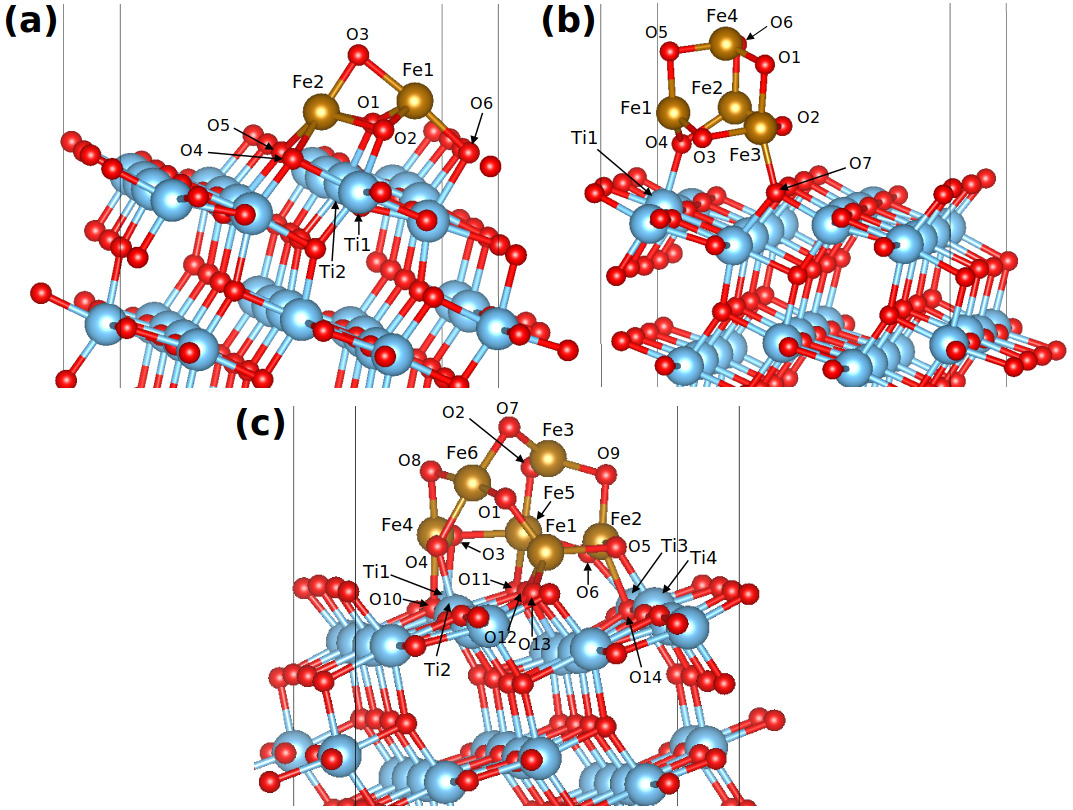}
\caption{Optimized structures of (a) (\ch{Fe2O3})$_1$/\ch{TiO2}, (b) (\ch{Fe2O3})$_2$/\ch{TiO2} and (c) (\ch{Fe2O3})$_3$/\ch{TiO2}. Selected atoms are labeled in the figure.}
\label{fig:heterostructures}
\end{figure}
\subsection{Structural and electronic properties of the (\ch{Fe2O3})$_n$/\ch{TiO2} heterostructures}

\subsubsection{Structural optimization}
The optimized structures of the (\ch{Fe2O3})$_n$/\ch{TiO2} heterostructures are shown in figure \ref{fig:heterostructures}. Importantly, the results showed that \ch{Fe2O3} clusters remained stable on the \ch{TiO2} surface without breaking down. For the (\ch{Fe2O3})$_1$ the optimization yielded a closed cage-like structure on the surface that resembles a pyramid with a parallelogram shaped base. The (\ch{Fe2O3})$_2$ and (\ch{Fe2O3})$_2$ maintained a cage-like structure at the \ch{TiO2} surface after the adsorption. The coordination number of the (\ch{Fe2O3})$_1$ and (\ch{Fe2O3})$_3$ were found to be five and nine, respectively, while the coordination number of the (\ch{Fe2O3})$_2$ was only two. The adsorption of the clusters resulted in some lattice distortion at the \ch{TiO2} surface primarily limited to the top layer of the surface. The nearest Ti and O atoms at the surface are generally shifted toward the clusters to form new Ti-O and Fe-O bonds between the surface and clusters. The surface Ti - O distances were 1.88 \AA, 1.98 \AA\ and in the range of 1.94 to 2.12 \AA, and Fe - surface O distances in the range of 2.04 to 2.14 \AA, 1.95 \AA\ and in the range of 1.87 to 2.02 \AA\ in the (\ch{Fe2O3})$_1$/\ch{TiO2}, (\ch{Fe2O3})$_2$/\ch{TiO2} and (\ch{Fe2O3})$_3$/\ch{TiO2}, respectively. We also investigated the structure of the clusters after the adsorption. In the adsorbed (\ch{Fe2O3})$_1$ cluster the Fe-O bond lengths varied from 1.84 to 2.14 \AA, showing significant change in the bond lengths when compared with the freestanding (\ch{Fe2O3})$_1$ cluster. In the adsorbed (\ch{Fe2O3})$_2$ and (\ch{Fe2O3})$_3$ clusters the distances lied in the range of 1.69 to 1.91 \AA, and 1.69  to 1.94 \AA, respectively. These are similar to that of freestanding (\ch{Fe2O3})$_2$ and (\ch{Fe2O3})$_3$ clusters. Detailed information on structural parameters are listed in table S4. The energetic stability was checked by calculating the adsorption energies of the (\ch{Fe2O3})$_1$, (\ch{Fe2O3})$_1$ and (\ch{Fe2O3})$_3$. The energies were -2.28 eV, -1.72 eV and -3.24 eV, respectively, showing the cluster-surface interaction to be energetically favourable. More negative adsorption energy indicates stronger interaction with \ch{TiO2}, and which, in this case, correlates with the number of newly formed bonds between the clusters and \ch{TiO2}.

\subsubsection{Electronic structure analysis}
\begin{figure}\centering
\includegraphics[width=1\linewidth]{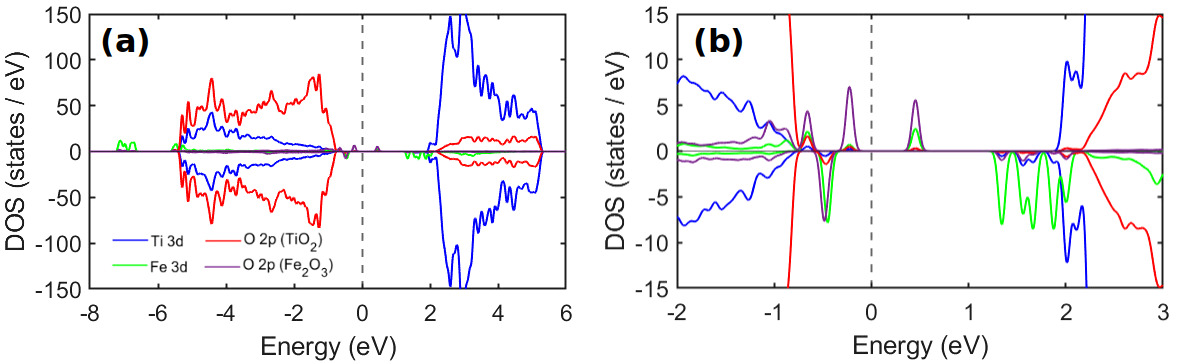}
\includegraphics[width=1\linewidth]{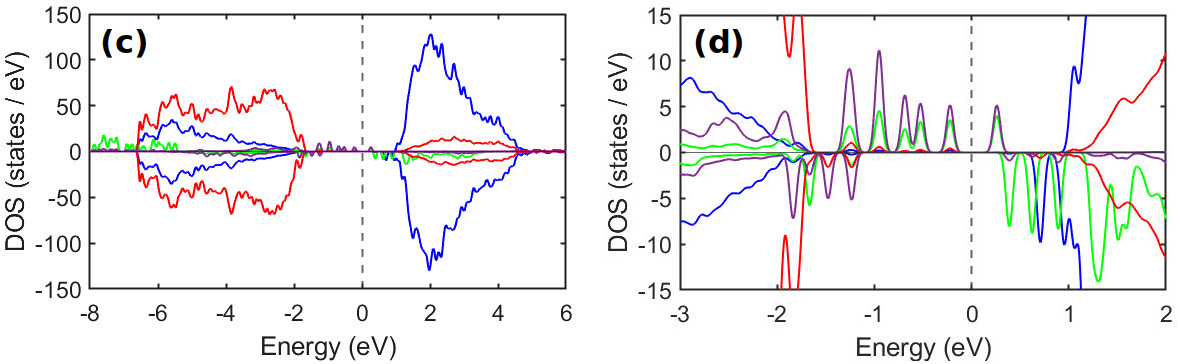}
\includegraphics[width=1\linewidth]{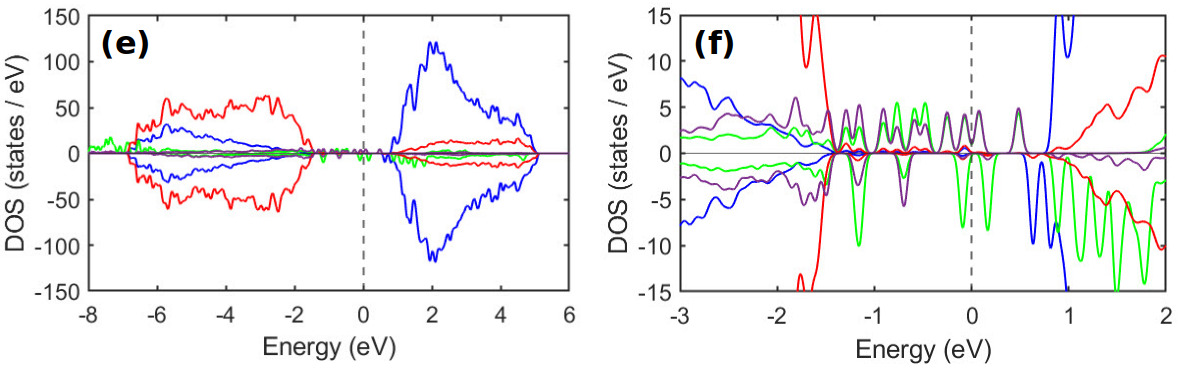}
\caption{The computed DOS of (\ch{Fe2O3})$_1$/\ch{TiO2} (a and b), (\ch{Fe2O3})$_2$/\ch{TiO2} (c and d) and (\ch{Fe2O3})$_3$/\ch{TiO2} (e and f). Projected DOSs and band gap regions are shown in the left and right, respectively. Fermi level is set to  zero energy.}
\label{fig:DOS-heterostructures}
\end{figure}

The electronic structure of the (\ch{Fe2O3})$_n$/\ch{TiO2} heterostructures are shown in figure \ref{fig:DOS-heterostructures}. The properties of primitive \ch{TiO2} remained unchanged but the results revealed an emergence of spin polarized, both occupied and unoccupied, \ch{Fe2O3}-states within the band gap of \ch{TiO2} due the cluster-surface interaction. The impurity states below the Fermi level led to an upward shift of valence band. The VBM was shifted from -0.05 eV of \ch{TiO2} to -0.70 eV in the (\ch{Fe2O3})$_1$/\ch{TiO2}, -1.60 eV in the (\ch{Fe2O3})$_2$/\ch{TiO2}, and -1.35 eV in the (\ch{Fe2O3})$_3$/\ch{TiO2}. These newly emerging electronic states resulted in a narrowing in band gap energy compared to the original band gap of \ch{TiO2}. The unoccupied states in the band gap divided the band gap of (\ch{Fe2O3})$_1$/\ch{TiO2} into two parts, 0.45\ eV and 0.90\ eV. In the (\ch{Fe2O3})$_2$/\ch{TiO2} the band gap was narrowed down to 0.20 eV. Adsorption of (\ch{Fe2O3})$_3$ induced states some of which cross the Fermi level, and thus (\ch{Fe2O3})$_3$/\ch{TiO2} exhibited a metallic nature. This is proposed to enable the activation of \ch{TiO2} by visible light. Due to the emerging states the VB of \ch{Fe2O3} locates above the VB of \ch{TiO2} and the CB of \ch{Fe2O3} below the CB of \ch{TiO2}. The particular band alignment (VB$_{\mathrm{\ch{TiO2}}} <$ VB$_{\mathrm{\ch{Fe2O3}}} <$ CB$_{\mathrm{\ch{Fe2O3}}} <$ CB$_{\mathrm{\ch{TiO2}}}$) could generally indicate a formation of type \rom{1} heterojunction with a straddling gap between \ch{Fe2O3} and \ch{TiO2}. This is in line with the study of Moniz \textit{et al.} \cite{Moniz}, and it can affect the charge transfer properties in the heterostructures. Our results showed that the \ch{Fe2O3} concentration has a significant impact on band gap energy. The decrease in it was more obvious in the presence of larger clusters. Mei \textit{et al.} \cite{Mei} have synthesized sheet-like \ch{TiO2}/\ch{Fe2O3} nanocomposites with different \ch{Fe2O3} content on \ch{TiO2} surface. Their results indicated improved light adsorption with \ch{Fe2O3} content, and by increasing it, the adsorption ability appeared to approach that of pure \ch{Fe2O3}, which showed the highest activity in the visible light region. Our results could imply that the \ch{Fe2O3} concentration is a determining factor in the photosensitization of \ch{TiO2} surface caused by the adsorbed \ch{Fe2O3} clusters. Conducting electronic states at the Fermi level induced by (\ch{Fe2O3})$_3$ could be advantageous for carrier transfer and contribute to improved photoresponse.

\subsubsection{Charge transfer analysis}
\begin{figure}[h!]\centering
\includegraphics[width=0.8\linewidth]{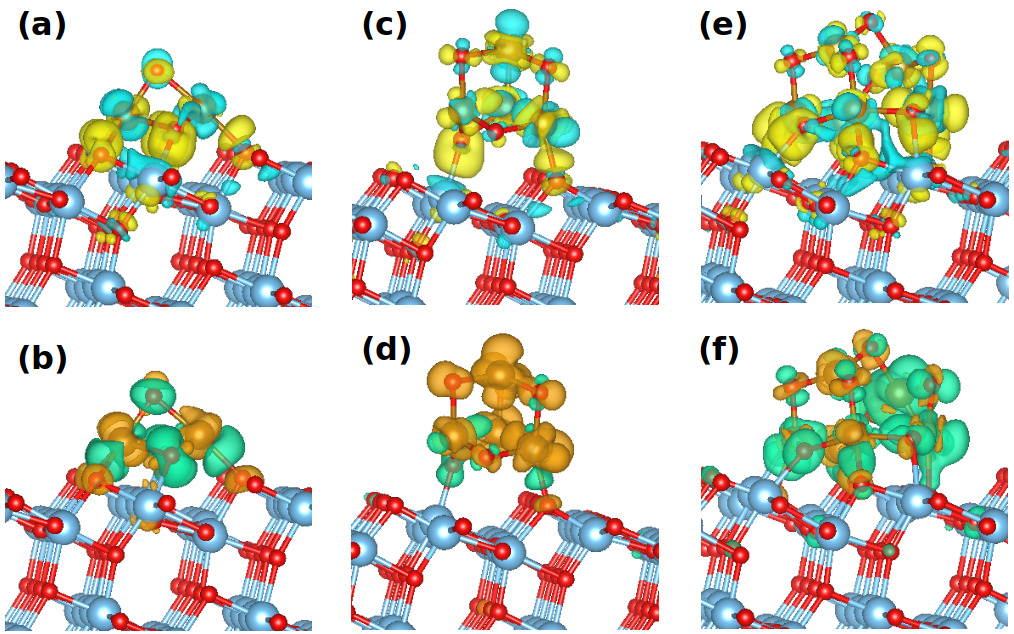}
\caption{The CDD and SDD plot of the (\ch{Fe2O3})$_1$/\ch{TiO2} (a and b),  (\ch{Fe2O3})$_2$/\ch{TiO2} (c and d) and (\ch{Fe2O3})$_3$/\ch{TiO2} (e and f). The labels of the atoms are provided in figure \ref{fig:heterostructures}.}
\label{fig:CDD-SDD-1heterostructure}
\end{figure}

Efficient charge separation and transfer are essential factors affecting the photocatalytic activity. To gain further insight into the charge transfer properties in the heterostructures we calculated the CDD, shown in figures \ref{fig:CDD-SDD-1heterostructure}, and performed the Bader analysis. The Bader charge and magnetic moment of selected atoms are listed in table S3. The results indicated a charge redistribution in the systems after the adsorption of the clusters, localized on the cluster and the top layer of the \ch{TiO2} surface. At the interface the electron accumulation and depletion regions were aligned in the direction of the formed bonds between \ch{TiO2} and \ch{Fe2O3}. This suggests that the bonds can be considered covalent with a polar feature. The blue isosurface around Fe showed charge depletion from Fe, indicating electron donor nature for the clusters. Increased amount of charge accumulation was observed around Fe atoms in case of larger \ch{Fe2O3} clusters. The work function of \ch{TiO2} was observed to be affected by the clusters. The work function decreased from 7.23 eV to 6.70 eV, 5.90 eV and 6.18 eV in the (\ch{Fe2O3})$_1$/\ch{TiO2}, (\ch{Fe2O3})$_2$/\ch{TiO2} and (\ch{Fe2O3})$_3$/\ch{TiO2}, respectively, supporting donor nature for \ch{Fe2O3} clusters. The lower work function can also facilitate electron injection from the \ch{TiO2} surface.

The Bader charge of the (\ch{Fe2O3})$_1$, (\ch{Fe2O3})$_2$ and (\ch{Fe2O3})$_3$ were 0.32 \textit{e}, 0.036 \textit{e} and 0.42 \textit{e}, suggesting an electron transfer from \ch{Fe2O3} clusters to \ch{TiO2}. Bader charges indicated an occurrence of notable charge transfer in the presence of (\ch{Fe2O3})$_1$ and (\ch{Fe2O3})$_3$ whereas the adsorption of the (\ch{Fe2O3})$_2$ induced only a small charge displacement. Nearly 90\% of the transferred charge was redistributed to the surface layer of \ch{TiO2}. Neither the CDD plots nor Bader charges showed charge localization at the surface, indicating even distribution of charge among surface layer atoms. It should be noted that our results are in conflict with Moniz \textit{et al.} \cite{Moniz} who proposed an electron transfer from \ch{TiO2} to \ch{Fe2O3} based on their theoretical study on the combined system of \ch{TiO2} and \ch{Fe2O3} clusters. The difference in the results may result from the use of different methods. Donor nature of Fe atoms was confirmed by the Bader charges. The Bader charges of the Fe1 and Fe2 atoms were 1.61 \textit{e} and 1.52 \textit{e} in the (\ch{Fe2O3})$_1$/\ch{TiO2}. The correlation between the Bader charges and oxidation state is still somewhat debuted. However, according to Posysaev \textit{et al.} \cite{55} the particular values suggested Fe$^{3+}$ oxidation state for Fe in the (\ch{Fe2O3})$_1$/\ch{TiO2}. Larger variation in the Bader charge of Fe atoms was observed in the adsorbed (\ch{Fe2O3})$_2$ and (\ch{Fe2O3})$_3$. In the (\ch{Fe2O3})$_2$/\ch{TiO2} the Bader charges of Fe1, Fe2, Fe3 ad Fe4 were 1.37 \textit{e}, 1.39 \textit{e}, 1.61 \textit{e} and 1.45 \textit{e} while in the (\ch{Fe2O3})$_3$/\ch{TiO2} the values were 1.63 \textit{e}, 1.51 \textit{e}, 1.32 \textit{e}, 1.20 \textit{e}, 1.56 \textit{e} and 1.30 \textit{e} for Fe1, Fe2, Fe3, Fe4, Fe5 and Fe6, respectively. Reduction in Bader charge could indicate an emergence of Fe$^{2+}$ species \cite{55} and therefore, existence of different valences, Fe$^{2+}$ and Fe$^{3+}$, in these heterostructures. Increased amount of charge accumulation indicated by the CDD plot can support the conclusion.

We were also interested in the magnetism brought by \ch{Fe2O3} in order to evaluate the oxidation state of Fe. In general, Fe introduces ferromagnetism in the heterostructures which is supported by the positive spin density around Fe in the SDD plots (figures \ref{fig:CDD-SDD-1heterostructure}). We found a total magnetic moment of $9.61\ \mathrm{\mu_B}$, $17.2\ \mathrm{\mu_B}$ and $17.8\ \mathrm{\mu_B}$ per unit cell for the (\ch{Fe2O3})$_1$/\ch{TiO2}, (\ch{Fe2O3})$_2$/\ch{TiO2} and (\ch{Fe2O3})$_3$/\ch{TiO2}. In the (\ch{Fe2O3})$_1$/\ch{TiO2} the magnetic moments of the Fe1 and Fe2 were $4.11\ \mathrm{\mu_B}$ and $4.06\ \mathrm{\mu_B}$, respectively, indicating high spin configuration and $3d^5$ occupation \cite{56,66}, that is, Fe$^{3+}$ oxidation state. In the (\ch{Fe2O3})$_2$/\ch{TiO2} the Fe atoms had an odd contribution to the total magnetic moment. We found a large magnetic moment of $4.02\ \mathrm{\mu_B}$ and $4.13\ \mathrm{\mu_B}$ for the Fe2 and Fe3, whereas a reduced magnetic moment of around $3.26\ \mathrm{\mu_B}$ and $3.36\ \mathrm{\mu_B}$ for the Fe1 and Fe4, respectively. In the adsorbed (\ch{Fe2O3})$_3$, Fe1, Fe2 and Fe5 exhibited a magnetic moment of 4.10\ $\mathrm{\mu_B}$, 3.67\ $\mathrm{\mu_B}$ and 3.96\ $\mathrm{\mu_B}$, respectively. The magnetic moment of Fe3 and Fe4 were only around 2.90\ $\mathrm{\mu_B}$, and Fe6 showed anti-parallel spin orientation with a magnetic moment of -0.842\ $\mathrm{\mu_B}$. The SDD plot of (\ch{Fe2O3})$_3$/\ch{TiO2} also showed an increased amount of spin down components in the cluster. Thus, ferrimagnetism occurred in the (\ch{Fe2O3})$_3$/\ch{TiO2}. With the varying Bader charges, the results can highly suggest mixed-valence state (Fe$^{2+}$/Fe$^{3+}$) for Fe in the (\ch{Fe2O3})$_2$/\ch{TiO2}. It is worth noticing that magnetic moments do not directly correlate with the Bader charges. Due to the charge transfer, spin magnetization was also transferred to the atoms nearby Fe. Bader charge and magnetic moment of selected atoms are listed in table S5.

Based on our results, the emerging \ch{Fe2O3} states in the band gap of \ch{TiO2} have an impact on the charge transfer mechanism in the heterostructures. Induced \ch{Fe2O3} states in the valence band are proposed to allow visible light excitation of electrons to the CB of \ch{TiO2}. Thus, charge accumulation occurs on the \ch{TiO2} surface. At the same time, the holes are found in the \ch{Fe2O3}. This could suggest spatial separation of electrons and hole, which can further imply reduction in recombination rate. In addition, the unoccupied \ch{Fe2O3}-states can act as traps for the conduction band electrons, changing the oxidation state of Fe to Fe$^{2+}$, hindering the charge transfer. However, because Fe$^{2+}$ is relatively unstable and the unoccupied Fe states lie close to the CB of \ch{TiO2}, Fe$^{2+}$ might tend to release the trap electrons and change back to Fe$^{3+}$ in the (\ch{Fe2O3})$_1$/\ch{TiO2}. In the (\ch{Fe2O3})$_2$/\ch{TiO2} the charge transfer between \ch{Fe2O3} and \ch{TiO2} was almost negligible, which could indicate less efficient charge separation in the system. A type \rom{1} band alignment with a small band gap energy might facilitate the recombination phenomenon occurring in \ch{Fe2O3}. In the (\ch{Fe2O3})$_2$/\ch{TiO2}, the highest charge transfer could be attributed to the \ch{Fe2O3} states crossing the Fermi level, enabling direct electron conduction to the CB. The intervalence charge transfer between Fe atoms with different oxidation states can also help to separate the charge in \ch{Fe2O3}.

\subsubsection{Hydrogen evolution reaction activity}

\begin{figure}[h!]\centering
\includegraphics[width=0.8\linewidth]{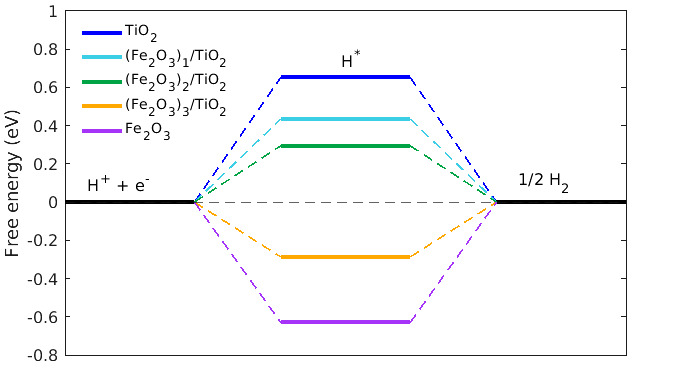} 
\caption{Free energy diagram for hydrogen evolution reaction on the pristine \ch{TiO2} and \ch{Fe2O3} surfaces and (\ch{Fe2O3})$_n$/\ch{TiO2} heterostructures.}
\label{fig:HER}
\end{figure}

After investigating the properties of the heterostructures we evaluate the HER activity by attaching a hydrogen atom on the heterostructures (figures S4) and calculated  the adsorption energy (figure \ref{fig:HER}). An ideal photocatalyst for the HER should have adsorption energy close to zero. Moreover, we incorporated a \ch{Fe2O3}-surface in our investigations for a better comparison. The adsorption energy in the pristine \ch{TiO2} and \ch{Fe2O3} surfaces were calculated to be 0.65 eV and -0.64 eV. A positive value indicates too weak interaction between \ch{TiO2} and hydrogen, inhibiting the adsorption. On the contrary, the adsorption of hydrogen on the \ch{Fe2O3} surface is exothermic with negative adsorption energy which facilitates the adsorption but makes the desorption of hydrogen difficult. The presence of the (\ch{Fe2O3})$_n$ clusters improved the adsorption energy, bringing it closer to the optimal zero energy. In the presence of (\ch{Fe2O3})$_1$ and (\ch{Fe2O3})$_2$ the adsorption energy decreased from 0.65 eV to 0.43 eV and 0.29 eV, respectively, showing weakened binding. In the (\ch{Fe2O3})$_3$/\ch{TiO2} the adsorption was much stronger with an energy value of -0.29. Thus, the \ch{Fe2O3} clusters facilitate the adsorption of hydrogen and can enhance the HER activity of \ch{TiO2}. Moreover, the larger cluster size enhanced the HER activity more than the smallest \ch{Fe2O3} cluster. This was attributed to the co-existence of Fe$^{3+}$ and Fe$^{2+}$ oxidation states for Fe. Interestingly, Zhang \textit{et al.} \cite{H2-mixed-valence}  reported similar findings on iron-based metal-organic frameworks, highlighting the mixed-valence of Fe in promoting the hydrogen evolution. Furthermore, these  results are in line with that of Sun \textit{et al.}, where they demonstrated that \ch{Fe2O3} clusters with small size and low coverage on \ch{TiO2} can be beneficial for catalytic reaction.


\subsection{Oxygen defect in the (\ch{Fe2O3})$_1$/\ch{TiO2} heterostructure}
\begin{figure}[h!]\centering
\includegraphics[width=1\linewidth]{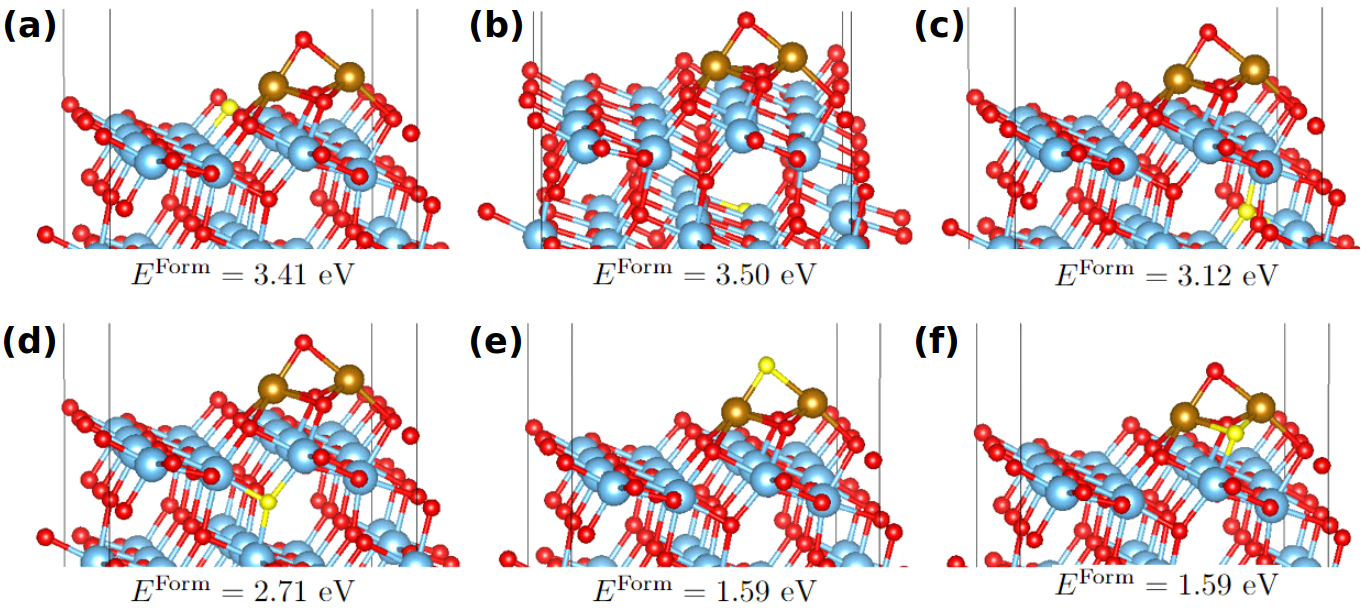}
\caption{The modeled defect sites in the (\ch{Fe2O3})$_1$/\ch{TiO2} and their formation energies, $E^{\mathrm{Form}}$. (a) Surface vacancy O$_\mathrm{v}$ and subsurface vacancies (b) O$_\mathrm{sv1}$, (c) O$_\mathrm{sv2}$, and (d) O$_\mathrm{sv3}$. The defects (e) O$_\mathrm{c1}$ and (f) O$_\mathrm{c2}$ are located in the (\ch{Fe2O3})$_1$ cluster. The defect sites are coloured yellow.}
\label{fig:sites-O-vacs}
\end{figure}

The modeled defect sites and their formation energies are shown in figure \ref{fig:sites-O-vacs}. Positive formation energies showed that the defect formation is endothermic. The formation energy of the O$_\mathrm{c1}$ and O$_\mathrm{c2}$ was the lowest, only 1.59 eV, indicating that the removal of oxygen is easier from the cluster than from the \ch{TiO2} surface. This can be explained by the weaker bonding of oxygen in the adsorbed cluster than at the \ch{TiO2} surface. Considering the relatively high formation energy of 3.41 eV of the surface vacancy O$_{\mathrm{v}}$, we decided to introduce several subsurface vacancy sites. Based on our findings, a subsurface vacancy is generally more stable in the (\ch{Fe2O3})$_1$/\ch{TiO2} than a surface vacancy. The  O$_\mathrm{sv3}$ was found to be the most stable, located underneath the cluster, with a formation energy of 2.71 eV. The formation energy of the O$_\mathrm{sv1}$ and O$_\mathrm{sv2}$ were 0.41 eV and 0.79 eV larger than that of O$_\mathrm{sv3}$, and their transverse distance, perpendicular to the $b$-axis, from the cluster were around 11 \AA \ and 4 \AA, respectively. This suggests that the cluster introduces a dependence of the formation energy on the distance from the cluster in the subsurface layer. Previously, Hoh \textit{et al}. \cite{58} have reported a distance dependence of oxygen vacancy formation energy from the Au-clusters on hematite surface, proposing that the cluster stabilizes the region underneath and near to it, making it more stable against reduction. We also observed the cluster to enhance the oxygen defect formation at the surface. In the pristine \ch{TiO2} surface the formation energies of the O$_\mathrm{v}$, O$_\mathrm{sv1}$, O$_\mathrm{sv2}$ and O$_\mathrm{sv3}$ were 5.47 eV, 5.46 eV, 5.12 eV, 5.46 eV, showing that the presence of the cluster can decrease the formation energy even by 50\%. A similar observation has been reported in previous studies \cite{59, 60}. 

Among the modeled defect sites, we chose to further investigate the effect of the O$_\mathrm{v}$, O$_\mathrm{sv3}$, and O$_\mathrm{c1}$ and O$_\mathrm{c2}$ on the properties of the (\ch{Fe2O3})$_1$/\ch{TiO2} heterostructure. We employed the same methods to investigate these defective heterostructures as for the defect-free heterostructures.


\subsubsection{Structural optimization}
\begin{figure}[h!]\centering
\includegraphics[width=0.9\linewidth]{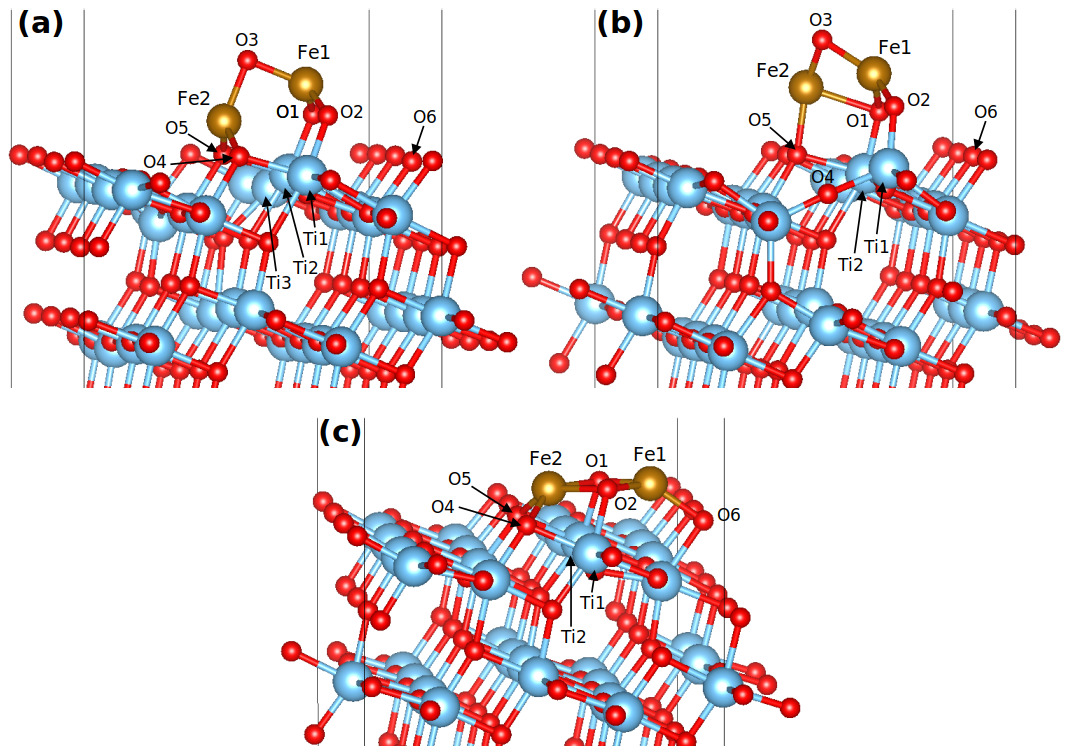}
\caption{The optimized structures of the defective heterostructures: (a) (\ch{Fe2O3})$_1$/\ch{TiO2}-O$_\mathrm{v}$, (b) (\ch{Fe2O3})$_1$/\ch{TiO2}-O$_\mathrm{sv3}$ and (c) (\ch{Fe2O3})$_1$/\ch{TiO2}-O$_\mathrm{c}$. The labels correspond to the same atoms in figure \ref{fig:heterostructures}a.}
\label{fig:Structures-O-vacs}
\end{figure}

The optimized structures of the defective heterostructures are shown in figure \ref{fig:Structures-O-vacs} while the bond distances between selected atoms are provided in table S6. It is evident that the defects affected the geometry of the cluster adsorbed on the surface. In the presence of the O$_\mathrm{v}$ and O$_\mathrm{sv3}$ the structure of the cluster became more open. The structural changes led to the breaking of bonds with \ch{TiO2}, thereby reducing the coordination number of  the (\ch{Fe2O3})$_1$ to four in the (\ch{Fe2O3})$_1$/\ch{TiO2}-O$_\mathrm{v}$ and three in the (\ch{Fe2O3})$_1$/\ch{TiO2}-O$_\mathrm{sv3}$. Furthermore, the O$_\mathrm{v}$ and O$_\mathrm{sv3}$ caused local lattice distortion at the \ch{TiO2} surface. The vacancies shifted the adjacent Ti atoms away from the vacancy site while the adjacent oxygen atoms moved closer to it. This is consistent with previous observations \cite{20}. The changes caused by the O$_\mathrm{v}$ were primarily limited to the surface layer whereas the O$_\mathrm{sv3}$ led to distortion notably extended to the first sublayer. The removal of either of the O$_\mathrm{c1}$ or O$_\mathrm{c2}$ resulted in the same geometry with the same total energy for the optimized systems. Therefore, we simply denoted the defective heterostructures simply as (\ch{Fe2O3})$_1$/\ch{TiO2}-O$_\mathrm{c}$. The remaining four atoms of the (\ch{Fe2O3})$_1$ were re-positioned in a manner that the top of the cluster became planar at the \ch{TiO2} surface, preserving the coordination number of five. 

\subsubsection{Electronic structure analysis}
\begin{figure}[h!]\centering
\includegraphics[width=1\linewidth]{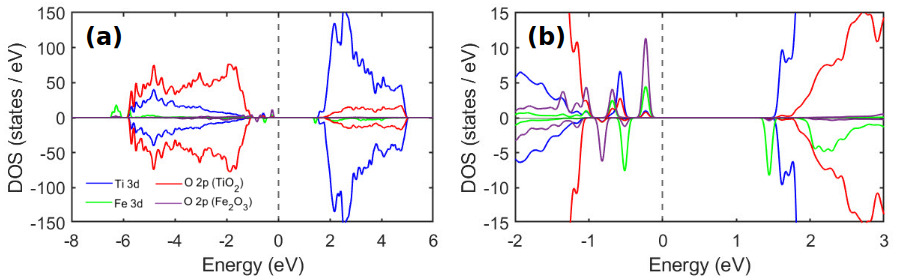}
\includegraphics[width=1\linewidth]{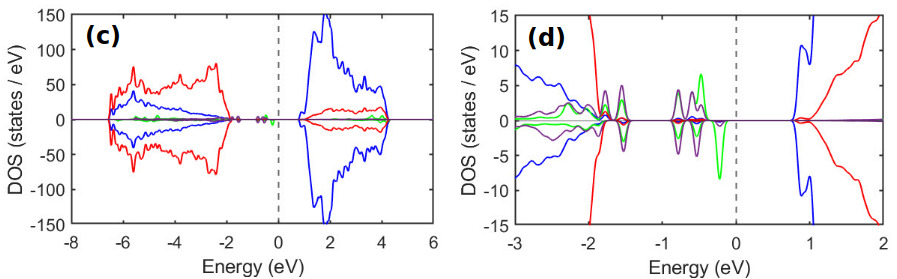}
\includegraphics[width=1\linewidth]{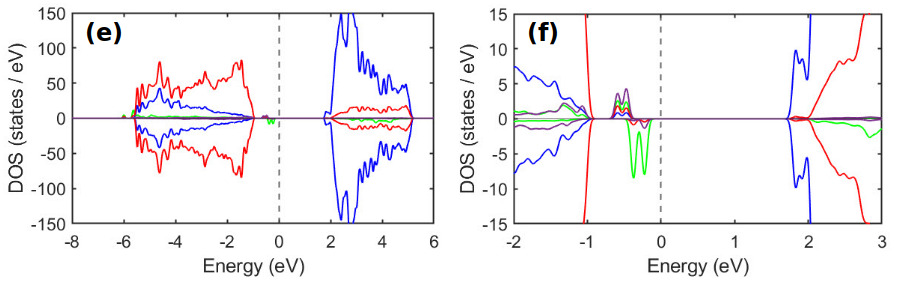}
\caption{The computed DOS of (a) (\ch{Fe2O3})$_1$/\ch{TiO2}-O$_\mathrm{v}$, (b) (\ch{Fe2O3})$_1$/\ch{TiO2}-O$_\mathrm{sv3}$ and (c) (\ch{Fe2O3})$_1$/\ch{TiO2}-O$_\mathrm{c}$. Fermi level is set to zero energy.}
\label{fig:DOS-O-vacs}
\end{figure}
Figure \ref{fig:DOS-O-vacs} represents the calculated DOS of the (\ch{Fe2O3})$_1$/\ch{TiO2}-O$_\mathrm{v}$, (\ch{Fe2O3})$_1$/\ch{TiO2}-O$_\mathrm{sv3}$ and  (\ch{Fe2O3})$_1$ /\ch{TiO2}-O$_\mathrm{c}$, showing changes in the band gap region. The valence band was further upward-shifted due to the oxygen defects. 
The VBM  positions of \ch{TiO2} were found to be at -0.96 eV in the (\ch{Fe2O3})$_1$/\ch{TiO2}-O$_\mathrm{v}$, -1.8 eV in the (\ch{Fe2O3})$_1$/\ch{TiO2}-O$_\mathrm{sv3}$, and -0.93 eV in the (\ch{Fe2O3})$_1$/\ch{TiO2}-O$_\mathrm{c}$, respectively. The properties of primitive \ch{TiO2} remained still unchanged. The presence of defects altered the band gap of the heterostructure. The (\ch{Fe2O3})$_1$/\ch{TiO2}-O$_\mathrm{sv3}$ showed a narrow band gap of 0.75 eV. Inversely, the O$_\mathrm{v}$ and O$_\mathrm{c}$ resulted in an increase in the band gap energy compared to the defect-free (\ch{Fe2O3})$_1$/\ch{TiO2}. We found a band gap of 1.45 eV for the (\ch{Fe2O3})$_1$/\ch{TiO2}-O$_\mathrm{v}$ and 1.80 eV for the (\ch{Fe2O3})$_1$/\ch{TiO2}-O$_\mathrm{c}$. Nevertheless, all the three defective heterostructures possessed a narrower band gap than the pristine \ch{TiO2} surface. Interestingly, a localized Ti 3d peak was observed at -0.57 eV, arising from the Ti3 atom (figure \ref{fig:Structures-O-vacs}). This can be evidence of formation of Ti$^{3+}$ species at the \ch{TiO2} surface, which is common phenomenon in \ch{TiO2} in the presence of oxygen vacancies.

\subsubsection{Charge density analysis}
\begin{figure}[h!]\centering
\includegraphics[width=0.9\linewidth]{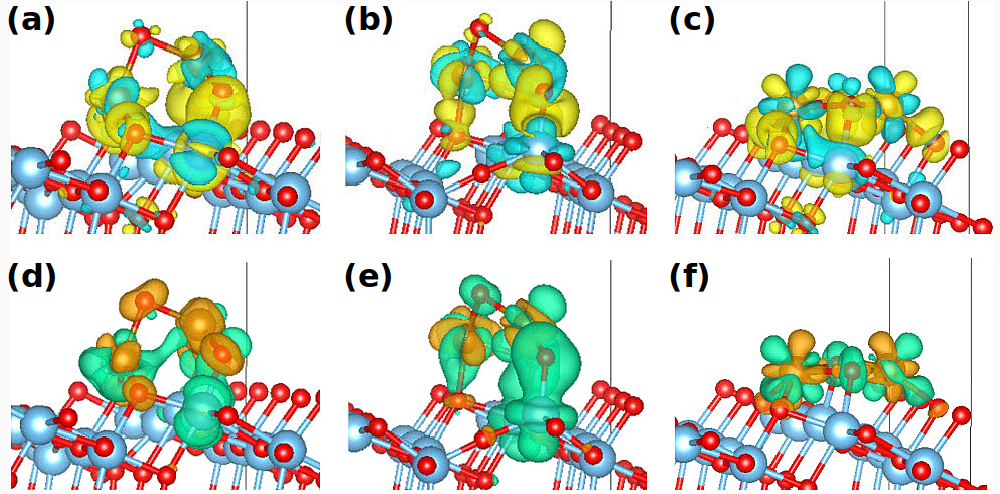}
\caption{The CDD (a, b, and c) and SDD (d, e, and f) of the (\ch{Fe2O3})$_1$/\ch{TiO2}-O$_\mathrm{v}$, (\ch{Fe2O3})$_1$/\ch{TiO2}-O$_\mathrm{sv3}$ and (\ch{Fe2O3})$_1$/\ch{TiO2}-O$_\mathrm{c}$.}
\label{fig:CDD-SDD-O-vacs}
\end{figure}

As previously, we investigated the changes in the charge transfer properties via the CDD and Bader analysis, supported by the SDD. The Bader charges and magnetic moments are listed in table S7 for selected atoms. The CDD plots, shown in figure (figures \ref{fig:CDD-SDD-O-vacs}), showed that the main charge redistribution occurred in the cluster and at the interface in the defective heterostructures as well. When a neutral oxygen defect is formed it releases effective excess charge which can generally be transferred either to Fe or Ti. The Bader analysis confirmed the charge gain of Fe, with a yellow isosurface observed around Fe atoms in the CDD plots. The calculated Bader charge of the Fe1 and Fe2 atoms were 1.52 \textit{e} and 1.30 \textit{e} in the (\ch{Fe2O3})$_1$/\ch{TiO2}-O$_\mathrm{v}$, 1.26 \textit{e} and 1.19 \textit{e} in the (\ch{Fe2O3})$_1$/\ch{TiO2}-O$_\mathrm{sv3}$, and 1.29 \textit{e} and 1.32 \textit{e} in the (\ch{Fe2O3})$_1$/\ch{TiO2}-O$_\mathrm{c}$, respectively. 
The corresponding magnetic moment of the Fe1 and Fe2 were 4.02 $\mathrm{\mu_B}$ and 3.77 $\mathrm{\mu_B}$ in the (\ch{Fe2O3})$_1$/\ch{TiO2}-O$_\mathrm{v}$, 3.26 $\mathrm{\mu_B}$ and -3.57 $\mathrm{\mu_B}$ in the (\ch{Fe2O3})$_1$/\ch{TiO2}-O$_\mathrm{sv3}$, and 3.67 $\mathrm{\mu_B}$ and 3.70 $\mathrm{\mu_B}$ in the  (\ch{Fe2O3})$_1$/\ch{TiO2}-O$_\mathrm{c}$, respectively. A lower magnetic moment supports a charge gain for Fe. To confirm lower spin magnetization of some Fe atoms, we calculated the SDD which showed a larger probability for spin down density (turquoise isosurface) around Fe (figure \ref{fig:CDD-SDD-O-vacs}). In the (\ch{Fe2O3})$_1$/\ch{TiO2}-O$_\mathrm{v}$ the total magnetic moment was still $9.66\ \mathrm{\mu_B}$ per unit cell whereas in the (\ch{Fe2O3})$_1$/\ch{TiO2}-O$_\mathrm{c}$ the total magnetic moment decreased to $7.79\ \mathrm{\mu_B}$. Surprisingly, the Fe2 atom in the (\ch{Fe2O3})$_1$/\ch{TiO2}-O$_\mathrm{sv3}$ experienced a magnetic phase transition. This resulted in a total magnetic moment of 0.09 $\mathrm{\mu_B}$. Due to the anti-parallel but unequal magnetic moment of Fe atoms, the (\ch{Fe2O3})$_1$/\ch{TiO2}-O$_\mathrm{sv3}$ exhibits weak ferrimagnetic behaviour. The magnetic phase transition was attributed to be to the notable lattice distortion at the \ch{TiO2} surface caused by the O$_\mathrm{sv3}$. For instance, in their study, Men$\mathrm{\acute{e}}$ndez \textit{et al.} \cite{67} have reported that the lattice distortion caused by oxygen defect can correlate with magnetic behavior and induce a magnetic phase transition.

Our findings suggest that Fe can have the mixed valence state (Fe$^{2+}$/Fe$^{3+}$) in the (\ch{Fe2O3})$_1$/\ch{TiO2}-O$_\mathrm{v}$ \cite{55, 56, 66}. We also observed localized spin density between the Fe atoms (figure \ref{fig:CDD-SDD-O-vacs}d). This can indicate magnetic coupling between the cations with different oxidation states, supporting the mixed-valence state for Fe. In the (\ch{Fe2O3})$_1$/\ch{TiO2}-O$_\mathrm{sv3}$ and (\ch{Fe2O3})$_1$/\ch{TiO2}-O$_\mathrm{c}$ the results indicates the Fe$^{2+}$ oxidation state for Fe \cite{55, 56, 66}. This was supported by the lower Bader charges and even contribution of Fe atoms to the spin magnetization. Additionally, in the (\ch{Fe2O3})$_1$/\ch{TiO2}-O$_\mathrm{v}$ we found the Bader charge of the Ti3 to change notably. It decreased from 2.04 \textit{e} to 1.79 \textit{e}, corresponding to Ti$^{3+}$ species \cite{55}. This is consistent with the findings from the DOS.

As per the Bader charge analysis the direction of the charge transfer remained from the cluster to the surface in the (\ch{Fe2O3})$_1$/\ch{TiO2}-O$_\mathrm{c}$. 0.45 \textit{e} of charge per unit cell was transferred to the surface, of which approximately 90\% \textit{e} was evenly distributed among Ti and O atoms of the surface layer. The charge transfer was almost 1.5 times larger than in the (\ch{Fe2O3})$_1$/\ch{TiO2}, showing an increase in the conductivity of \ch{TiO2}. This was attributed to the formation of a type \rom{2} heterojunction. Since no \ch{Fe2O3}-states exist below the CB of \ch{TiO2} electrons from the VB of \ch{TiO2} and \ch{Fe2O3} can be excited directly to the CB of \ch{TiO2} without trapping. Electron excitation to the unoccupied \ch{Fe2O3}-states can be possible but due to the type \rom{2} heterojunction electrons can be transferred to the CB of \ch{TiO2}. On the contrary, we observed that the direction of the charge transfer reverse in the presence of the $_\mathrm{v}$ and O$_\mathrm{sv3}$. 0.12 \textit{e} and 0.62 \textit{e} per unit cell were transferred from the surface to the cluster in the (\ch{Fe2O3})$_1$/\ch{TiO2}-O$_\mathrm{v}$ and (\ch{Fe2O3})$_1$/\ch{TiO2}-O$_\mathrm{sv3}$, respectively. A larger charge transfer indicates stronger interaction between \ch{TiO2} and \ch{Fe2O3} in the (\ch{Fe2O3})$_1$/\ch{TiO2}-O$_\mathrm{sv3}$, which is consistent with the lower formation energy of the O$_\mathrm{sv3}$ compared to that of O$_\mathrm{v}$. For the (\ch{Fe2O3})$_1$/\ch{TiO2}-O$_\mathrm{sv3}$ the Bader analysis can support the Z-scheme mechanism, leading to electron accumulation to \ch{Fe2O3}. The lowest unoccupied \ch{Fe2O3} states started to appear approximately 1.5 eV above the Fermi level, enabling electron excitation using visible light.

The oxygen defects further decreased the work function of the (\ch{Fe2O3})$_1$/\ch{TiO2}. A neutral oxygen defect is a positively charged which makes negatively charged surface more positive, that is, decrease the work function. In the presence of the O$_\mathrm{v}$, O$_\mathrm{sv3}$ and O$_\mathrm{c}$ we found the work function to be 5.27 eV, 5.28 eV and 5.41 eV, respectively. This suggests that an oxygen defect can facilitate electron migration from the \ch{TiO2} surface. 

\subsubsection{Effect on hydrogen evolution} 
The results show that oxygen defects enable the engineering of electronic properties of the (\ch{Fe2O3})$_1$/
\ch{TiO2}. Finally, we evaluate the HER activity of the defective heterostructures. H-containing structures are shown in figure S4. Based on the results, oxygen defects have a detrimental effect on the HER activity of the (\ch{Fe2O3})$_1$/\ch{TiO2}. The O$_\mathrm{v}$, O$_\mathrm{sv3}$ and O$_\mathrm{c}$ increased the adsorption energy of hydrogen from 0.43 eV to 0.71 eV, 0.61 eV and 0.60 eV, respectively. Thus, the interaction with hydrogen becomes weaker again, indicating decrease in HER activity. The HER activity of the defective heterostructures is comparable with the pristine \ch{TiO2} surface. 

\section{Conclusions}
First principles DFT calculations were employed to investigate the heterostructure of the \ch{TiO2} surface and (\ch{Fe2O3})$_n$ ($n=1,2,3$) clusters, and superior catalytic performance was indicated for the heterostructures. The adsorption of the clusters on the surface was exothermic, thus enabling the modification of the surface properties of \ch{TiO2}. The formation of a heterojunction between \ch{TiO2} and \ch{Fe2O3} was identified to affect band gap energy, charge transfer and charge separation in the heterostructures, leading to enhanced photocatalytic properties. The clusters were observed to improve the HER activity of \ch{TiO2}. The existence of mixed-valence Fe (Fe$^{2+}$/Fe$^{3+}$) led to greater enhancement in the HER activity. We also incorporated oxygen defects on the (\ch{Fe2O3})$_1$/\ch{TiO2}, and the defect formation was found to be endothermic. The presence of the (\ch{Fe2O3})$_1$ cluster substantially enhanced the oxygen defect formation at \ch{TiO2} compared to the pristine surface. We found that electronic structure and charge transfer and the local magnetism can be tuned by oxygen defects, which can indirectly affect the oxidation state of Fe. However, it was observed that an oxygen defect was not beneficial in order to improve the HER activity. 

Our investigation also suggests variations of the dye-sensitized solar cells originally proposed by Grätzel \cite{G}. Here, hematite nanoparticles play the role of dyes used to sensitize \ch{TiO2} in order to produce \ch{H2}/\ch{O2} by water photoelectrolysis. The hematite nanoparticles are bound to the nanostructured wide-band-gap \ch{TiO2} semiconductor and can be used as photosensitizers to harvest solar energy and generate excitons needed to fulfill the requirements for a photovoltaic scheme. 

In summary, the results of our work reveal three important points. The first point is that +2 and +3 valence-state mixing offers a lot of potential to control the electronic structure of hematite clusters. The second point is that further electronic structure optimization could be achieved by engineering these clusters with various oxygen defects. The third point is that the presence of Fe$^{2+}$-Fe$^{3+}$ intervalence charge transfer implies charge transfer between these non-equivalent Fe sites and could result in broad absorption in the visible or IR region of the electromagnetic spectrum.

\section*{\large Acknowledgement}
This work is supported by the Finnish cultural foundation (Grant No. 00230235) and the Kvantum Institute. The authors acknowledge the CSC - IT Center for Science, Finland, for computational resources.

\section*{\large Declaration of competing interest}
The authors declare no competing financial interests.


\begin{thebibliography}{0}

\bibitem{1}Villa K, Gal$\mathrm{\acute{a}}$n-Mascar$\mathrm{\acute{o}}$s J R, L$\mathrm{\acute{o}}$pez N,  Palomares E2021 \textit{Sustain. Energy Fuels} \textbf{5}, 4560.
\bibitem{2}Fujishima A, Honda K 1972 \textit{Nature} \textbf{238}, 37.
\bibitem{3}Nosaka Y, Nakamura M, Hirakawa T 2002 \textit{Phys. Chem. Chem. Phys}. \textbf{4}, 1088.
\bibitem{4}Chen X, Mao S S 2007 \textit{Chem. Rev}. \textbf{107}, 2891.
\bibitem{5}Fujishima A, Zhang X, Tryk D A 2008 \textit{ Surf. Sci. Rep}. \textbf{63}, 515.
\bibitem{G}Grätzel M 2001 \textit{Nature} \textbf{414}, 338-344.
\bibitem{6}Bahnemann D W, Hilgendorff M, Memming R 1997 \textit{J. Phys. Chem. B} \textbf{101}, 4265.
\bibitem{R}Li R, Y. Weng Y, Zhou X, Wang X, Mi Y, Chong R,Han H, Li C 2015 \textit{Energy Environ. Sci.} \textbf{8}, 2377.

\bibitem{brookite}Di Paola A, Bellardita M, Palmisano L 2013 \textit{Catalysts} \textbf{3}, 36. 
\bibitem{A1}Sclafani A, Herrmann J M 1996 \textit{J. Phys. Chem.} \textbf{100}, 13655.
\bibitem{A2}Luttrell T, Halpegamage S, Tao J, Kramer A, Sutter E, Batzill M 2014 \textit{Sci. Rep.} \textbf{4}, 4043.
\bibitem{A3}$\mathrm{\check{Z}}$erjav G, $\mathrm{\check{Z}}$i$\mathrm{\check{z}}$ek K, Zava$\mathrm{\check{s}}$nik, Pintar A 2022 \textit{J. Environ. Chem. Eng.} \textbf{10}, 107722.
\bibitem{A4}Augustynski J 1993 \textit{Electrochim. Acta} \textbf{38}, 43.
\bibitem{A5}Hirano M, Nakahara C, Ota K, Tanaike O, Inagaki M 2003 \textit{J. Solid State Chem.} \textbf{170}, 39. 
\bibitem{A6}Zhang J, Zhou P, Liu J, Yu J 2014 \textit{Phys. Chem. Chem. Phys.} \textbf{16}, 20382.
\bibitem{R1}Ohno T, Sarukawa K, Matsumura M 2001 \textit{Phys. Chem. B} \textbf{105}, 2417.
\bibitem{R2}Ohno T, Tokieda K, Higashida S, Matsumura M 2003 \textit{Appl. Catal. A} \textbf{244}, 383.
\bibitem{M1}Liao Y, Que W 2010 \textit{J. Alloys Compd.} \textbf{505}, 243.

\bibitem{7}Wang H, Zhang L, Chen Z, Hu J, Li S, Wang Z, Liu J, Wang X 2014 \textit{Chem. Soc. Rev}. \textbf{43}, 5234.
\bibitem{11}Muraro P C L, Mortar S R, Vizzotto B S, G. Chuy G, Dos Santos C, Brum L F W, da Silva W L, 2020 \textit{Sci. Rep.} \textbf{10}, 3055.
\bibitem{12}Glasscock J A, Barnes P R F, Plumb I C, Bendavid A, Martin P J 2008 \textit{Thin Solid Films} \textbf{516}, 1716.
\bibitem{Mei} Mei Q, Zhang F, Wang N, Yang Y, Wu R, Wang W 2019 \textit{RSC Adv.} \textbf{9}, 22764.
\bibitem{Singh}Singh A P, Wang R B, Tossi C,  Tittonen I, Wickman B, Hellman A 2021 \textit{RSC Adv.} \textbf{11}, 4297.
\bibitem{14}Cao Y Q, Zi T Q, Zhao X R, Liu C, Ren Q, Fang J B, Li W M, Li A D, 2020 \textit{Sci. Rep.} \textbf{10}, 13437. 
\bibitem{8}Lee S C, Lintang H O, Yuliati L 2017 \textit{Beilstein J. Nanotechnol}. \textbf{8}, 915. 
\bibitem{15}Sun Q, Leng W, Li Z, Xu Y 2012 \textit{J. Hazard. Mater.} \textbf{229-230}, 224. 
\bibitem{plasma}Bootluck W, Chittrakarn T,  Techato K, Khongnakorn W 2021 \textit{J. Environ. Chem. Eng.} \textbf{9}, 105660.
\bibitem{bacteria}Sharma B, Boruah P K, Yadav A, Das M R 2018 \textit{J. Environ. Chem. Eng.} \textbf{6}, 134.
\bibitem{Abbas}Abbas N, Shao G N, Haider M S, Imran S M, Park S S, Kim H T 2016 \textit{J. Ind. Eng. Chem.} \textbf{39}, 112.
\bibitem{17}Deng J, Zhuo Q, Lv X 2019 \textit{J. Electroanal. Chem.} \textbf{835}, 287. 
\bibitem{13}Xia Y, Yin L 2013 \textit{Phys. Chem. Chem. Phys.} \textbf{15}, 18627. 
\bibitem{9}Li X, Lin H, Chen X, Niu H, Liu J,  Zhang T, Qu F 2016 \textit{Phys. Chem. Chem. Phys}. \textbf{18}, 9176. 



\bibitem{19}Pacchioni G 2003 \textit{ChemPhysChem} \textbf{4}, 1041.
\bibitem{23}Pan X, Yang M Q, Fu X, Zhang N, Xu Y J 2013 \textit{Nanoscale} \textbf{5}, 3601.
\bibitem{25}Elahifard M, Sadrian M R, Mirzanejad A, Behjatmanesh-Ardakani R, Ahmadvand S 2020 \textit{Catalysts} \textbf{10}, 397.
\bibitem{26}Scheiber P, Fidler M, Dulub O, Schmid M, Diebold U, Hou W, Aschauer U, Selloni A 2005 \textit{Surf. Sci.} \textbf{598}, 226. 
\bibitem{C1}Trenczek-Zajac A, Synowiec M, Zakrzewska K, Zazakowny K, Kowalski K, Dziedzic A, Rabecka M 2022 \textit{ACS Appl. Mater. Interfaces} \textbf{14}, 33.
\bibitem{C2}Synowiec M, Z$\mathrm{\Acute{a}}$kutn$\mathrm{\Acute{a}}$ D, Trenczek-Zajac A, Rabecka M 2023 \textit{Appl. Surf. Sci.} \textbf{608}, 155186.
\bibitem{C3}Wu Q, Zheng Q, van de Krol R 2012 \textit{J. Phys. Chem. C} \textbf{12}, 7219.
\bibitem{C4}Zhu J, Ren J, Huo Y, Bian Z, Li H 2007 \textit{J. Phys. Chem. C} \textbf{111}, 18965.

\bibitem{31}Ahmed A Y, Kandiel T A, Oekermann T, Bahnemann D 2011 \textit{J. Phys. Chem. Lett.} \textbf{2}, 2461.
\bibitem{30}G$\mathrm{\ddot{u}}$nnemann C, Haisch C, Fleisch M, Schneider J, Emeline A V, Bahnemann D W 2019 \textit{ACS Catal.} \textbf{9}, 1001.

\bibitem{34}Kresse G, Hafner J 1993 \textit{Phys. Rev. B} \textbf{47}, 558.
\bibitem{35}Kresse G, Furthm$\mathrm{\ddot{u}}$ller J 1996 \textit{Comput. Mater. Sci.} \textbf{6}, 15.
\bibitem{36}Kresse G, Furthm$\mathrm{\ddot{u}}$ller J 1996 \textit{Phys. Rev. B} \textbf{54}, 11169.
\bibitem{37}Perdew J P, Burke K, Ernzerhof M 1996 \textit{Phys. Rev. Lett.} \textbf{77}, 3865.
\bibitem{38}Dudarev S L, Botton G A, Savrasov S Y, Humphreys C J, Sutton A P 1998 \textit{Phys. Rev. B: Condens. Matter Mater. Phys.} \textbf{57}, 1505.
\bibitem{39}Kumaravel V, Rhatigan S, Mathew S,  Bartlett J, Nolan M, Hinder S J, Sharma P K, Singh A, Bryne J A, Harrison J, Pillai S C 2019 \textit{J. Phys. Chem. C} \textbf{123}, 21083.
\bibitem{40}Rollmann G, Rohrbach A, Entel P, Hafner J, 2004 \textit{Phys. Rev. B} \textbf{69}, 165107.
\bibitem{UO}Ataei S S, Mohammadizadeh M R, Seriani N 2016 \textit{J. Phys. Chem. C} \textbf{120}, 8421.

\bibitem{42}Monkhorst H J, Pack J D 1976 \textit{Phys. Rev. B} \textbf{13}, 5188.
\bibitem{41}Bl{\"o}chl P E 1994 \textit{Phys. Rev. B} \textbf{50}, 17953.
\bibitem{43}Momma K, Izumi F, 2008 \textit{J. Appl. Cryst.} \textbf{41}, 653. 
\bibitem{vaspkit}Wang V, Xu N, Liu J C , Tang G, Geng W T 2021 \textit{Comput. Phys. Commun.} \textbf{267}, 108033. 
\bibitem{45}Tang W, Sanville E, Henkelman G, 2009 \textit{J. Phys.: Condens. Matter} \textbf{21}, 084204.
\bibitem{46}Sanville E, Kenny S D, Smith R, Henkelman G 2007 \textit{J. Comp. Chem.} \textbf{28}, 899. 
\bibitem{47}Henkelman G, Arnaldsson A, J$\mathrm{\acute{o}}$nsson H 2006 \textit{Comput. Mater. Sci.} \textbf{36}, 354.
\bibitem{48}Yu M, Trinkle D R, 2011 \textit{J. Chem. Phys.} \textbf{134}, 064111.

\bibitem{Fe2O3-parameters}Finger W, Hazen R M 1980 \textit{J. Appl. Phys.} \textbf{51}, 5362.
\bibitem{44}Burdett J K, Hughbanks T, Miller G J, Richardson Jr. J W, Smith J V 1987 \textit{J. Am. Chem. Soc.} \textbf{109}, 3639.
\bibitem{I1}Morgan B J, Watson G W 2007 \textit{Surf. Sci.} \textbf{601}, 5034.
\bibitem{I2}Mattioli G, Alippi P, Filippone F, Caminiti, Bonapasta A A 2010 \textit{J. Phys. Chem. C} \textbf{114}, 21694.
\bibitem{I3}M. Cococcioni, S. de Gironcoli, Phys. Rev. B 71 (2005), 035105.
\bibitem{I4}Erhart P, Albe K, Klein A 2006 \textit{Phys. Rev. B} \textbf{73}, 205203.
\bibitem{I5}Ghosal S, Dutta K, Chowdhury S, Jana D 2022 \textit{J. Phys.D: Appl. Phys.} \textbf{55}, 375303.
\bibitem{Labat}Labat F, Baranek P, Domain C, Minot C, Adamo C 2007 \textit{J. Chem. Phys.} \textbf{126}, 154703.
\bibitem{DiValentin}Di Valentin C, Pacchioni G, Selloni A 2004 \textit{Phys. Rev. B} \textbf{70}, 085116.
\bibitem{Fe2O3-DOS1}Rivera R, Pinto H P, Stashans A, Piedra L, 2012 \textit{Phys. Scr.} \textbf{85}, 015602.
\bibitem{Fe2O3-DOS2}Huda M N, Walsh A, Yan Y, Wei S H, Al-Jassim M M 2010 \textit{J. Appl. Phys.} 107, 123712.
\bibitem{Fe2O3-DOS3}Fujimori A, Saeki M, Kimizuka N, Taniguchi M, Suga S 1986 \textit{Phys. Rev. B} \textbf{34}, 7318.
\bibitem{P1}Mo S D, Ching W Y 1994 \textit{Phys. Rev. B}, \textbf{51}, 13023.
\bibitem{HSE2}Yamamoto T, Ohno T 2012 \textit{Phys. Chem. Chem. Phys.} \textbf{14}, 589.
\bibitem{HSE1}Meng Y, Liu X W, Huo C F, Guo W P, Cao D B, Peng Q, Dearden A, Gonze X, Yang Y, Wang, J, Jiao H, Li Y, Wen X D 2016 \textit{J. Chem. Theory Comput.} \textbf{12}, 5132.
\bibitem{HSE3}De$\Acute{\mathrm{a}}$k P, Aradi B, Frauenheim T 2011 \textit{Phys. Rev. B} \textbf{83}, 155207.
\bibitem{Moniz}Moniz S J A, Shevlin S A, An X, Guo Z X, Tang J 2014 \textit{Chem. Eur. J.} \textbf{20}, 15571.
\bibitem{51}Erlebach A, H$\mathrm{\ddot{u}}$hn C, Jana R, Sierka M 2014  \textit{Phys. Chem. Chem. Phys.} \textbf{6}, 26461. 
\bibitem{Cluster1}Peters L, Sasioglu E, Rossen S, Friedrich C,  Bl$\mathrm{\ddot{u}}$gel S, Katsnelson M I 2017 \textit{Phys. Rev. B} \textbf{95}, 155119. 
\bibitem{Cluster2}Reilly N M, Reveles J U, Johnson G E, del Campo M, Khanna S N, Ko A M, Castleman A W 2007 \textit{J. Phys. Chem. C} \textbf{111}, 19086. 
\bibitem{TM-FexOy}Majid A, Zahid S, Khan S U D, Khan, S U D 2020 \textit{J. Nanopart. Res.} \textbf{22}, 145.
\bibitem{55}Posysaev S, Miroshnichenko O, Alatalo M, Le D, Rahman T S 2019  \textit{Comput. Mater. Sci.} \textbf{161}, 403.
\bibitem{52}Wang J, Huang J, Meng J, Li Q, Yang, J 2017 \textit{RSC Adv.} \textbf{7}, 39877.
\bibitem{54}Kashiwaya S, Morasch J, Streibel V, Toupance T, Jaegermann W, Klein A 2018 \textit{Surface} \textbf{1}, 73.
\bibitem{WF}Borodin A, Reichling M 2011 \textit{Phys. Chem. Chem. Phys.} \textbf{13}, 15442.
\bibitem{56}Neufeld O, Toroker M C 2015 \textit{J. Phys. Chem. C} \textbf{119}, 5836.
\bibitem{66}Geneste G, Paillard C, Dkhil B 2019 \textit{Phys. Rev. B} \textbf{99}, 024104.
\bibitem{H2-mixed-valence}Zhang X, Ma X, Ye Y, Guo C, Xu X, Zhou J, Wang B 2023 \textit{Chem. Eng. J.} \textbf{456}, 140939.

\bibitem{58}Hoh S W, Thomas L, Jones G, Willock D J 2015 \textit{Res. Chem. Intermed.} \textbf{41}, 9587.
\bibitem{59}Giordano L, Goniakowski J, Pacchioni G 2001 \textit{Phys. Rev. B} \textbf{64}, 075417.
\bibitem{60}Yang Z, Lu Z, Luo G, Hermansson K 2007 \textit{Phys. Lett. A} \textbf{369}, 132.
\bibitem{20}Li H, Guo Y, Robertson J 2015 \textit{J. Chem. Phys. C} \textbf{119}, 18160.



\bibitem{67}Men$\mathrm{\acute{e}}$ndez C, Chu D, Carzola C 2020 \textit{Comput. Mater.} \textbf{6}, 76.

\end{thebibliography}
\end{document}


\title{\Large\textbf{Supplementary information} \\ \vspace{0.3cm}
\Large Understanding and Optimizing the Sensitization of Anatase Titanium Dioxide Surface with Hematite Clusters}
\author{Kati Asikainen\ $^{1*}$, Matti Alatalo $^1$, Marko Huttula $^1$, \\
B. Barbiellini $^2$ and S. Assa Aravindh\ $^{1*}$ \\[0.3cm] \normalsize ${}^1$\textit{Nano and Molecular Systems Research Unit, University of Oulu, FI-90014, Finland} \\[0.2cm] \normalsize ${}^2$\textit{Lappeenranta-Lahti University of Technology (LUT), FI-53851 Lappeenranta, Finland}\\
\begin{tabular}{l}
\\[0.1cm]
\small ${}^*$Corresponding author: Kati.Asikainen@oulu.fi \\
\small ${}^*$Corresponding author: Assa.SasikalaDevi@oulu.fi
\end{tabular}
}

\maketitle

\beginsupplement
\clearpage

\begin{table}[h!]
\caption{The effect of the Hubbard U correction on lattice parameters and band gap energy of bulk anatase \ch{TiO2}.}
\begin{tabular}{llll}
\textbf{U values}                  &  & \textbf{Lattice parameters (\AA)} & \textbf{Band gap (eV)} \\ \hline
\rule{0pt}{\normalbaselineskip}\textbf{U$_{\mathrm{eff}}$(Ti)=3.0eV, U$_{\mathrm{eff}}$(O)=0 eV } &  & \quad $a=3.88, b=9.71$         & \quad \quad 2.1 \\[0.1cm]
\textbf{U$_{\mathrm{eff}}$(Ti)=4.5eV, U$_{\mathrm{eff}}$(O)=0 eV}    && \quad $a=3.91, b=9.73$  & \quad \quad 2.3           \\[0.1cm]
\textbf{U$_{\mathrm{eff}}$(Ti)=6.0eV, U$_{\mathrm{eff}}$(O)=0 eV}    && \quad $a=3.94, b=9.78$  & \quad \quad 2.6           \\[0.1cm]
\textbf{U$_{\mathrm{eff}}$(Ti)=7.0eV, U$_{\mathrm{eff}}$(O)=0 eV}    && \quad $a=3.97, b=9.81$  & \quad \quad 2.9           \\[0.1cm]
\textbf{U$_{\mathrm{eff}}$(Ti)=4.5eV, U$_{\mathrm{eff}}$(O)=5.25 eV} && \quad $a=3.91, b=9.73$  & \quad \quad 2.4  \\[0.1cm] \hline        
\end{tabular}
\end{table}

\noindent Several different U values were tested for \ch{TiO2} by applying the U correction only for Ti 3d states or both Ti 3d and O 2p states. The results are listed in table S1. Using the GGA, we obtained lattice parameters of  $a=3.81\ \mathrm{\AA}$\ and $b=9.75\ \mathrm{\AA}$, and a band gap of 1.7 eV. Our results showed that increasing the U$_{\mathrm{eff}}$(Ti), without applying the U correction for O 2p states, provided a more realistic band gap closer to the experimentally measured optical band gap, but simultaneously, it led to a larger structural deviation compared to the GGA. We also concluded that U$_{\mathrm{eff}}$(Ti)=4.5 eV we chose for our calculations improved the band gap a reasonable amount without causing too significant deviation in the lattice parameters. To test the effect of U$_{\mathrm{eff}}$(O), we applied it with U$_{\mathrm{eff}}$(Ti)=4.5 eV. For U$_{\mathrm{eff}}$(O) we used a value of 5.25 eV according to Yamamoto \textit{et al.} \cite{HSE2}. The results showed that applying U$_{\mathrm{eff}}$(O) with U$_{\mathrm{eff}}$(Ti) did not affect the lattice parameters, and provided only a small improvement in the description of the band gap. It is worth noticing that Yamamoto \textit{et al.} \cite{HSE2} obtained a band gap of 2.72 eV using U(Ti)=4.2 eV and U(O)=5.25 eV, which is fairly consistent with our results. Furthermore, the primary disagreement with the band gap obtained by Ataei \textit{et al.} \cite{UO} can be explained by the utilization of different pseudopotentials. Specifically, we employ the projected augmented wave (PAW) method, whereas they use the Troullier-Martins (TM) approach. Notably, the PAW is generally acknowledged for its superior accuracy compared to the TM.



\begin{figure}[h!]\centering
\includegraphics[width=0.91\linewidth]{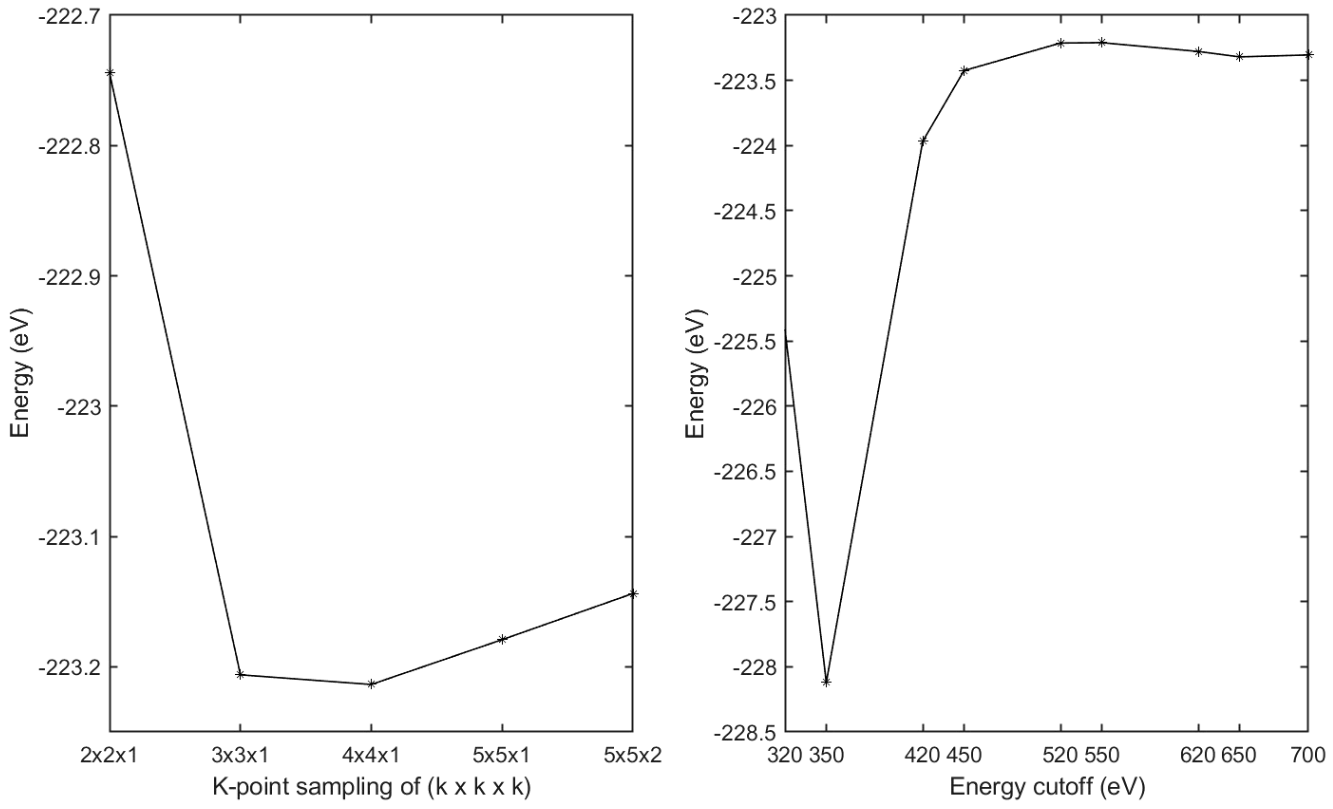} 
\caption{The convergence of the total energy with respect to k-point sampling (on the left) and energy cutoff (on the right) was performed for the bulk \ch{Fe2O3}, the unit cell of which is shown in figure S5.\\[0.2cm] \indent Based on the results the optimal energy cutoff and k-point sampling were found to be 650\ eV and $4 \times 4 \times 1$. The chosen energy cutoff was employed for the rest of the calculations. For modeling the bulk \ch{Fe2O3}, a k-point sampling of $4\times 4\times 1$ was used.}
\label{fig:convergence_tests}
\end{figure}

\clearpage



\begin{figure}[h!]\centering
\includegraphics[width=0.65\linewidth]{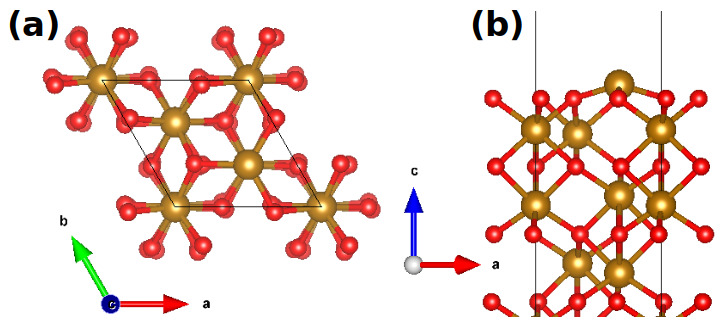}
\caption{(a) Top view and (b) side view of the \ch{Fe2O3}(0001) surface with Fe-O$_3$-Fe-R termination. \\[0.2cm]
Besides the pristine \ch{TiO2} surface, we also included the \ch{Fe2O3} surface into our investigations. According to previous studies (0001)-surface is the most stable and dominant facet in natural and synthesized \ch{Fe2O3}, and thus intensively studied \cite{Fe1,Fe2,Fe3}. Pristine \ch{Fe2O3}(0001) has three possible surface terminations: Fe-O$_3$-Fe-R (single iron layer), Fe-Fe-O$_3$-R (double iron layer), and O$_3$-Fe-Fe-R (oxygen layer) \cite{Fe1}. Of the different terminations, Fe-O$_3$-Fe-R has been reported to be energetically the most favorable at room temperature and over the wide range of oxygen chemical potential \cite{Fe2, Fe3, FeC1}, and therefore, we decided to model the \ch{Fe2O3}(0001) surface with a single Fe layer (a and b). The (0001) surface was created by taking the bulk structure of \ch{Fe2O3} (figure S4a) and adding a vacuum with a thickness of around 20 Å along the z-direction. After the relaxation, the optimized structure with ($2 \times 2 \times 1$) supercell of \ch{Fe2O3}(0001) surface was taken in order to investigate the hydrogen adsorption.}
\label{fig:Fe2O3(0001)}
\end{figure}

\clearpage

\begin{figure}[h!]\centering
\includegraphics[width=0.7\linewidth]{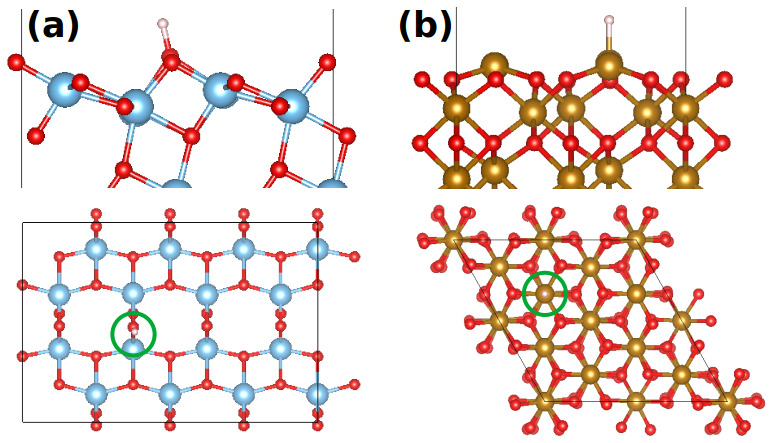}
\caption{Hydrogen adsorption on the (a) \ch{TiO2} and \ch{Fe2O3} surfaces (side and top view). Adsorption site is highlighted with a green circle. 
}
\label{fig:H-attached}
\end{figure}


\begin{figure}[h!]\centering
\includegraphics[width=0.71\linewidth]{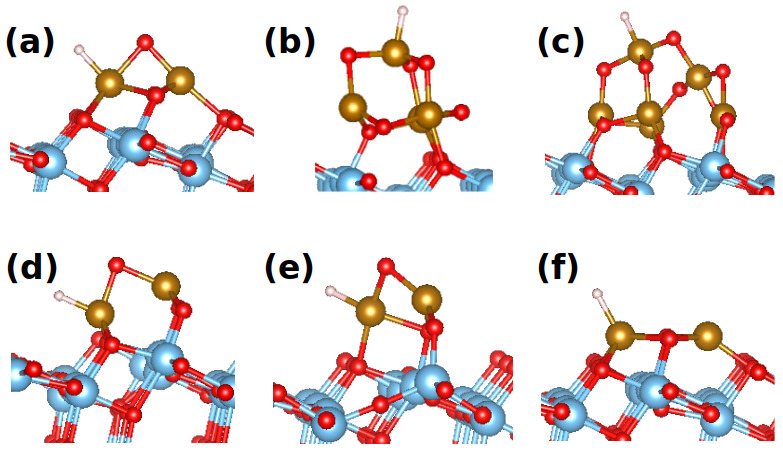}
\caption{The relaxed structures of the defect-free and reduced heterostructures after the adsorption of hydrogen (H is marked as white atom): (a) (\ch{Fe2O3})$_1$/\ch{TiO2}, (b) (\ch{Fe2O3})$_2$/\ch{TiO2}, (c) (\ch{Fe2O3})$_3$/\ch{TiO2}, (d)(\ch{Fe2O3})$_1$/\ch{TiO2}-O$_{\mathrm{v}}$, (e) (\ch{Fe2O3})$_1$/\ch{TiO2}-O$_{\mathrm{sv3}}$ and (f) (\ch{Fe2O3})$_1$/\ch{TiO2}-O$_{\mathrm{c}}$.}
\label{fig:H-attached}
\end{figure}

\begin{figure}[h!]\centering
\includegraphics[width=0.48\linewidth]{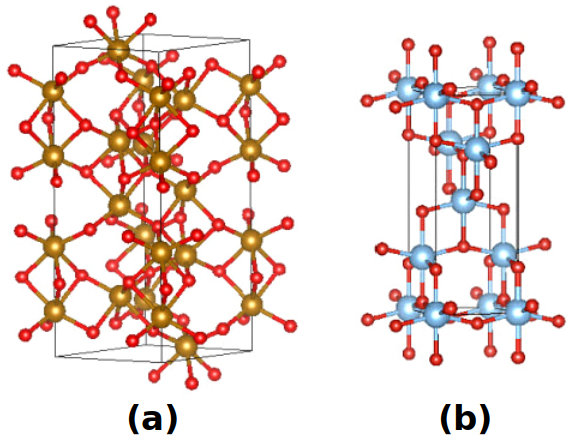}
\caption{Schematic illustration of the crystal structure of the (a) bulk \ch{Fe2O3} and (b) tetragonal anatase \ch{TiO2}. We used the hexagonal representation for the unit cell of \ch{Fe2O3}, containing 30 atoms in total (12 Fe and 18 O atoms), and tetragonal unit cell for anatase \ch{TiO2}, containing 12 atoms in total (4 Ti and 8 O atoms) \cite{1}.}
\label{fig:bulk-Fe2O3-TiO2}
\end{figure}

\vspace{1cm}

\begin{figure}[h!]\centering
\includegraphics[width=1\linewidth]{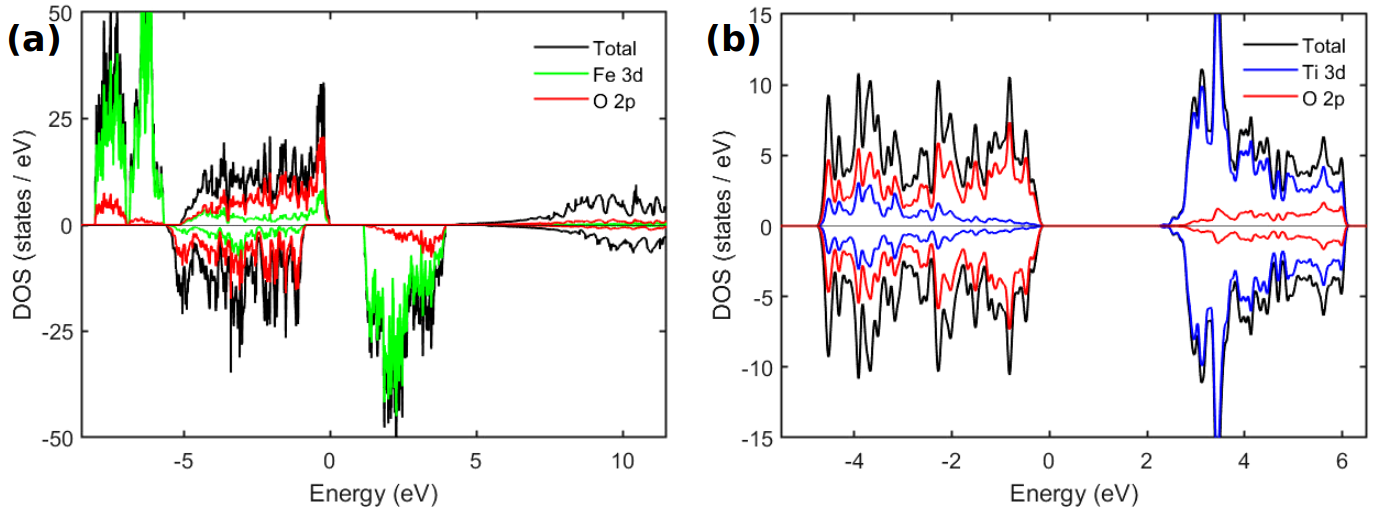}
\caption{The computed density of states of the (a) bulk \ch{Fe2O3} and (b) bulk \ch{TiO2} using the GGA+U functional. \\[0.2cm]
To calculate the electronic structure, we used the k-point sampling of $9\times 9\times 4$ for bulk \ch{TiO2}. To obtain better quality for the DOS of hematite we increased the k-point sampling symmetrically from $4\times 4\times 1$ to $8\times 8\times 3$ and smearing width from 0.05\ eV to 0.1\ eV. Using the GGA functional \ch{Fe2O3} was predicted to be metallic and for \ch{TiO2} we found a band gap of 1.7\ eV. After employing the Hubbard correction we found a band gap of 1.16\ eV for \ch{Fe2O3} and 2.3\ eV for \ch{TiO2}. Besides, we found a magnetic moment of $4.328\ \mathrm{\mu_B}$ for Fe in the \ch{Fe2O3} which is also comparable to the experimental value of $4.9\ \mathrm{\mu_B}$ \cite{mag-mom}. \ch{TiO2} was correctly predicted to be non-magnetic.}
\label{fig:DOS-Fe2O3-TiO2}
\end{figure}

\begin{table}[h!]\centering
\caption{Fe-O bond lengths in (\ch{Fe2O3})$_1$, (\ch{Fe2O3})$_2$ and (\ch{Fe2O3})$_3$. The atom labeling is also shown below.}
\includegraphics[width=0.65\linewidth]{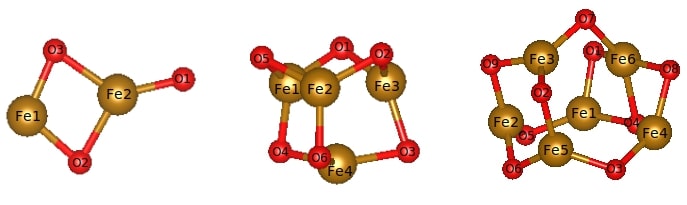}
\begin{tabular}{llll|lll|ll}
\hline \rule{0pt}{1.1\normalbaselineskip}& \multicolumn{2}{l}{\textbf{\quad\quad (\ch{Fe2O3})$_1$}} && \multicolumn{2}{l}{\textbf{\quad\quad (\ch{Fe2O3})$_2$}} & & \multicolumn{2}{l}{\textbf{\quad\quad (\ch{Fe2O3})$_3$}}\Tstrut \\[0.1cm]
\hline
 \rule{0pt}{1.5\normalbaselineskip}\makecell[l]{\textbf{Bond}\\\textbf{lengths (\AA)}}  & &  &  &   &  &  &  &   \\

\rule{0pt}{\normalbaselineskip}
&\quad \textbf{Fe1-O1} &\quad 1.70 &&\quad \textbf{Fe1-O1} &\quad 1.74 &&\quad \textbf{Fe1-O1} &\quad 1.76 \\[0.1cm]
&\quad \textbf{Fe1-O2} &\quad 1.70 &&\quad \textbf{Fe1-O4} &\quad 1.75 &&\quad \textbf{Fe1-O4} &\quad 1.83 \\[0.1cm]
&\quad \textbf{Fe2-O1} &\quad 1.87 &&\quad \textbf{Fe1-O5} &\quad 1.74 &&\quad \textbf{Fe1-O5} &\quad 1.77 \\[0.1cm]
&\quad \textbf{Fe2-O2} &\quad 1.87 &&\quad \textbf{Fe2-O2} &\quad 1.73 &&\quad \textbf{Fe2-O5} &\quad 1.81 \\[0.1cm]
&\quad \textbf{Fe2-O3} &\quad 1.65 &&\quad \textbf{Fe2-O5} &\quad 1.78 &&\quad \textbf{Fe2-O6} &\quad 1.76 \\[0.1cm]
& & &&\quad \textbf{Fe2-O6} &\quad 1.73 &&\quad \textbf{Fe2-O9} &\quad 1.82 \\[0.1cm]
& & &&\quad \textbf{Fe3-O1} &\quad 1.83 &&\quad \textbf{Fe3-O2} &\quad 1.69 \\[0.1cm]
& & &&\quad \textbf{Fe3-O2} &\quad 1.80 &&\quad \textbf{Fe3-O7} &\quad 1.71 \\[0.1cm]
& & &&\quad \textbf{Fe3-O3} &\quad 1.80 &&\quad \textbf{Fe3-O9} &\quad 1.71 \\[0.1cm]
& & &&\quad \textbf{Fe4-O3} &\quad 1.80 &&\quad \textbf{Fe4-O3} &\quad 1.77 \\[0.1cm]
& & &&\quad \textbf{Fe4-O4} &\quad 1.83 &&\quad \textbf{Fe4-O4} &\quad 1.92 \\[0.1cm]
& & &&\quad \textbf{Fe4-O6} &\quad 1.80 &&\quad \textbf{Fe4-O8} &\quad 1.81   \\[0.1cm]
&& & & & &&\quad  \textbf{Fe5-O2} &\quad 1.81 \\[0.1cm]   
&& & & & &&\quad  \textbf{Fe5-O3} &\quad 1.76 \\[0.1cm]   
&& & & & &&\quad  \textbf{Fe5-O6} &\quad 1.78 \\[0.1cm]   
&& & & & &&\quad  \textbf{Fe6-O1} &\quad 1.85 \\[0.1cm]   
&& & & & &&\quad  \textbf{Fe6-O4} &\quad 1.99 \\[0.1cm]   
&& & & & &&\quad  \textbf{Fe6-O7} &\quad 1.82 \\[0.1cm]   
&& & & & &&\quad  \textbf{Fe6-O8} &\quad 1.82 \\[0.3cm]  \hline  
\end{tabular}
\label{tab:O-vacs-bonds}
\end{table}

\begin{table}[h!]\centering
\caption{The Bader charges (in \textit{e}) and magnetic moments (in $\mathrm{\mu_B}$) of atoms in the (\ch{Fe2O3})$_1$, (\ch{Fe2O3})$_2$ and (\ch{Fe2O3})$_3$. The sum of spin magnetic moments was 9.28$\ \mathrm{\mu_B}$, 13.66 $\ \mathrm{\mu_B}$ and 22.90$\ \mathrm{\mu_B}$, respectively. The atom labeling is also shown above in table S2.}
\scalebox{1}{
\begin{tabular}{lll|lll|lll}
\hline    & \multicolumn{2}{l|}{\rule{0pt}{1.1\normalbaselineskip}\textbf{(\ch{Fe2O3})$_1$}} &     & \multicolumn{2}{l|}{\textbf{(\ch{Fe2O3})$_2$}} &     & \multicolumn{2}{l}{\textbf{(\ch{Fe2O3})$_3$}} \\[0.2cm] \hline
 & \makecell[l]{\rule{0pt}{1.1\normalbaselineskip}\textbf{Bader}\\\textbf{charge}}   & \makecell[l]{\rule{0pt}{1.1\normalbaselineskip}\textbf{Magnetic}\\\textbf{moment}}  &     & \makecell[l]{\rule{0pt}{1.1\normalbaselineskip}\textbf{Bader}\\\textbf{charge}} & \makecell[l]{\rule{0pt}{1.1\normalbaselineskip}\textbf{Magnetic}\\\textbf{moment}} &     & \makecell[l]{\rule{0pt}{1.1\normalbaselineskip}\textbf{Bader}\\\textbf{charge}} & \makecell[l]{\rule{0pt}{1.1\normalbaselineskip}\textbf{Magnetic}\\\textbf{moment}}\Bstrut\\
\hline
\rule{0pt}{1.1\normalbaselineskip}\textbf{Fe1} & \hspace{0.2cm}1.08 & \hspace{0.4cm}3.32  & \textbf{Fe1} & \hspace{0.2cm}1.24 & \hspace{0.4cm}2.85  & \textbf{Fe1} & \hspace{0.2cm}1.42  & \hspace{0.4cm}3.50 \\[0.1cm]
\textbf{Fe2} &  \hspace{0.2cm}1.46 & \hspace{0.4cm}3.60 & \textbf{Fe2} & \hspace{0.2cm}1.26 & \hspace{0.4cm}2.89 & \textbf{Fe2} & \hspace{0.2cm}1.19 & \hspace{0.4cm}3.65 \\[0.1cm]
\textbf{O1}&  \hspace{0.2cm}-0.714 & \hspace{0.4cm}0.421 & \textbf{Fe3} & \hspace{0.2cm}1.33 & \hspace{0.4cm}3.70 & \textbf{Fe3} & \hspace{0.2cm}1.03 & \hspace{0.4cm}3.78 \\[0.1cm]
\textbf{O2}&  \hspace{0.2cm}-0.908 &  \hspace{0.4cm}0.083 & \textbf{Fe4} & \hspace{0.2cm}1.32  & \hspace{0.4cm}3.66 & \textbf{Fe4}  & \hspace{0.2cm}1.46 & \hspace{0.4cm}3.94 \\[0.1cm]
\textbf{O3} &  \hspace{0.2cm}-0.912 &  \hspace{0.4cm}0.084 & \textbf{O1} & \hspace{0.2cm}-0.838 &  \hspace{0.4cm}0.051 & \textbf{Fe5}  & \hspace{0.2cm}1.18 & \hspace{0.4cm}2.97 \\[0.1cm]
&  &  & \textbf{O2} & \hspace{0.2cm}-0.979 & \hspace{0.4cm}0.126 & \textbf{Fe6}  & \hspace{0.2cm}1.53 & \hspace{0.4cm}4.04  \\[0.1cm]
&  &  & \textbf{O3} & \hspace{0.2cm}-0.980 & \hspace{0.4cm}0.271 & \textbf{O1} & \hspace{0.2cm}-0.718 & \hspace{0.4cm}-0.275 \\[0.1cm]
&  &  & \textbf{O4} & \hspace{0.2cm}-0.965 & \hspace{0.4cm}0.038 & \textbf{O2} & \hspace{0.2cm}-0.965 & \hspace{0.4cm}0.065 \\[0.1cm]
&  &  & \textbf{O5} & \hspace{0.2cm}-0.847 & \hspace{0.4cm}-0.044 & \textbf{O3} &\hspace{0.2cm}-0.914 & \hspace{0.4cm}0.222 \\[0.1cm]
&  &  & \textbf{O6} & \hspace{0.2cm}-0.807 & \hspace{0.4cm}0.121 & \textbf{O4} & \hspace{0.2cm}-0.749 & \hspace{0.4cm}0.074 \\[0.1cm]
&  &  &  &  &  & \textbf{O5} & \hspace{0.2cm}-0.825 & \hspace{0.4cm}0.273 \\[0.1cm]
&  &  &  &  &  & \textbf{O6} & \hspace{0.2cm}-0.822 & \hspace{0.4cm}-0.038 \\[0.1cm]
&  &  &  &  &  & \textbf{O7} & \hspace{0.2cm}-0.926 & \hspace{0.4cm}0.465 \\[0.1cm]
&  &  &  &  &  & \textbf{O8} & \hspace{0.2cm}-0.975 & \hspace{0.4cm}0.329 \\[0.1cm]
&  &  &  &  &  & \textbf{O9} & \hspace{0.2cm}-0.909 & \hspace{0.4cm}-0.100 \\[0.1cm]
\hline
\end{tabular}}
\end{table}

\clearpage

\begin{table}[h!]\centering
\caption{The bond lengths between the selected atoms in the (\ch{Fe2O3})$_1$/\ch{TiO2}, (\ch{Fe2O3})$_2$/\ch{TiO2} and (\ch{Fe2O3})$_3$/\ch{TiO2}. The labels of the atoms are shown in figure 3.}
\begin{tabular}{ll P{1.5cm}|lP{1.5cm}|lP{1.5cm}}
\hline\rule{0pt}{1.1\normalbaselineskip} & \multicolumn{2}{l|}{\hspace{0.4cm}\textbf{(\ch{Fe2O3})$_1$/\ch{TiO2}}}\Tstrut & \multicolumn{2}{l|}{\hspace{0.4cm}\textbf{(\ch{Fe2O3})$_2$/\ch{TiO2}}}\Tstrut & \multicolumn{2}{l}{\hspace{0.4cm}\textbf{(\ch{Fe2O3})$_3$/\ch{TiO2}}}\Tstrut \\[0.1cm]
\hline
 \rule{0pt}{1.5\normalbaselineskip}\makecell[l]{\textbf{Bond}\\\textbf{lengths (\AA)}}\Tstrut  & &    &   &    \\

\rule{0pt}{\normalbaselineskip}
&\quad\textbf{Fe1-O1} & 1.88  &\quad  \textbf{Fe1-O3}& 1.76  &\quad \textbf{Fe1-O1} & 1.80 \\[0.1cm]
&\quad\textbf{Fe1-O2} & 1.88  &\quad  \textbf{Fe1-O4}& 1.91  &\quad\textbf{Fe1-O5} & 1.93 \\[0.1cm]
&\quad\textbf{Fe1-O3} & 1.99  &\quad  \textbf{Fe1-O5}& 1.85  &\quad\textbf{Fe2-O5} & 1.90 \\[0.1cm]
&\quad\textbf{Fe1-O6} & 2.04  &\quad  \textbf{Fe2-O2}& 1.74  &\quad\textbf{Fe2-O6} & 1.89 \\[0.1cm]
&\quad\textbf{Fe2-O1} & 2.14  &\quad  \textbf{Fe2-O4}& 1.90  &\quad\textbf{Fe2-O9} & 1.79 \\[0.1cm]
&\quad\textbf{Fe2-O2} & 2.14  &\quad  \textbf{Fe2-O6}& 1.87  &\quad\textbf{Fe3-O2} & 1.71 \\[0.1cm]
&\quad\textbf{Fe2-O3} & 1.84  &\quad  \textbf{Fe3-O1}& 1.90  &\quad\textbf{Fe3-O7} & 1.69 \\[0.1cm]
&\quad\textbf{Fe2-O4} & 2.12  &\quad  \textbf{Fe3-O2}& 1.80  &\quad\textbf{Fe3-O9} & 1.70 \\[0.1cm]
&\quad\textbf{Fe2-O5} & 2.12  &\quad  \textbf{Fe3-O3}& 1.82  &\quad\textbf{Fe4-O3} & 1.82 \\[0.1cm]
&\quad\textbf{Ti1-O2} & 1.88  &\quad  \textbf{Fe4-O1}& 1.70  &\quad\textbf{Fe4-O4} & 1.85 \\[0.1cm] 
&\quad\textbf{Ti2-O1} & 1.88  & \quad \textbf{Fe4-O5}& 1.69  &\quad\textbf{Fe4-O8} & 1.75 \\[0.1cm]    
& & &\quad\textbf{Fe4-O1}& 1.69 &\quad\textbf{Fe5-O2}& 1.78 \\[0.1cm]   
& & &\quad\textbf{Ti1-O4}& 1.98 &\quad\textbf{Fe5-O3}& 1.94 \\[0.1cm]
& & &\quad\textbf{Fe3-O7}& 1.95 &\quad\textbf{Fe5-O6}& 1.94 \\[0.1cm] 
& & & & & \quad\textbf{Fe6-O1}& 1.81 \\[0.1cm]
& & & & & \quad\textbf{Fe6-O4}& 1.94 \\[0.1cm]
& & & & & \quad\textbf{Fe6-O7}& 1.83 \\[0.1cm]
& & & & & \quad\textbf{Fe6-O8}& 1.85 \\[0.1cm]
& & & & & \quad\textbf{Fe1-O12}& 2.02\\[0.1cm]
& & & & & \quad\textbf{Fe1-O13}& 1.98\\[0.1cm]
& & & & & \quad\textbf{Fe2-O14}& 1.99\\[0.1cm]
& & & & & \quad\textbf{Fe4-O10}& 1.98\\[0.1cm]
& & & & & \quad\textbf{Fe5-O11}& 1.87\\[0.1cm]
& & & & & \quad\textbf{Ti1-O3}& 2.12\\[0.1cm]
& & & & & \quad\textbf{Ti2-O4}& 1.94\\[0.1cm]
& & & & & \quad\textbf{Ti3-O6}& 2.00\\[0.1cm]
& & & & & \quad\textbf{Ti4-O5}& 1.96\\[0.3cm] \hline
\end{tabular}
\label{tab:heterostructures-bonds}
\end{table}

\vspace{1cm}

\begin{table}[h!]\centering
\caption{The Bader charges (in \textit{e}) and magnetic moments (in $\mathrm{\mu_B}$) of the (\ch{Fe2O3})$_1$/\ch{TiO2}, (\ch{Fe2O3})$_2$/\ch{TiO2} and (\ch{Fe2O3})$_3$/\ch{TiO2}. The labels of the atoms are shown in figure 3.}
\scalebox{1}{
\begin{tabular}{lll|lll|lll}
\hline  \rule{0pt}{1.1\normalbaselineskip}  & \multicolumn{2}{l|}{\hspace{0.65cm}\textbf{(\ch{Fe2O3})$_1$/\ch{TiO2}}} &  & \multicolumn{2}{l|}{\hspace{0.35cm}\textbf{(\ch{Fe2O3})$_2$/\ch{TiO2}}} &  & \multicolumn{2}{l}{\hspace{0.35cm}\textbf{(\ch{Fe2O3})$_3$/\ch{TiO2}}}\Tstrut\\[0.2cm] \hline
  \rule{0pt}{1.5\normalbaselineskip}  & \makecell[l]{\textbf{Bader}\\\textbf{charge}}  & \makecell[l]{\textbf{Magnetic}\\\textbf{moment}}  &    & \makecell[l]  {\textbf{Bader}\\\textbf{charge}} & \makecell[l]{\textbf{Magnetic}\\\textbf{moment}} & &\makecell[l]  {\textbf{Bader}\\\textbf{charge}} & \makecell[l]{\textbf{Magnetic}\\\textbf{moment}} \Bstrut\\
\hline
\rule{0pt}{\normalbaselineskip}\textbf{Fe1} &\quad 1.61 &\quad 4.11  &\textbf{Fe1} &\quad 1.37 &\quad 3.26  &\textbf{Fe1} & \quad 1.63&\quad 4.10\\[0.1cm]
\textbf{Fe2} &\quad 1.52  &\quad 4.06  & \textbf{Fe2} &\quad 1.39 &\quad 4.02  &\textbf{Fe2} &\quad 1.51 &\quad 3.67\\[0.1cm]
\textbf{O1}  &\quad -0.987  &\quad 0.287   &\textbf{Fe3} &\quad 1.61  &\quad 4.13  &\textbf{Fe3} &\quad 1.33 &\quad 2.59\\[0.1cm]
\textbf{O2}  &\quad -0.984 &\quad 0.287 & \textbf{Fe4} &\quad 1.45  &\quad 3.36  &\textbf{Fe4} &\quad 1.20 &\quad 2.98\\[0.1cm]
\textbf{O3}  &\quad -0.897 &\quad 0.519 & \textbf{O1}  &\quad -0.90 &\quad 0.191  &\textbf{Fe5} &\quad 1.56 &\quad 3.96\\[0.1cm]
\textbf{O4} &\quad -1.07 &\quad 0.034 & \textbf{O2}  &\quad -0.96  &\quad 0.480   &\textbf{Fe6} &\quad 1.30 &\quad -0.842\\[0.1cm]
\textbf{O5}  &\quad -1.07  &\quad 0.034 & \textbf{O3}  &\quad -1.01 &\quad 0.414  &\textbf{O1} &\quad -0.913 &\quad 0.223\\[0.1cm]
\textbf{O6}  &\quad -1.11 &\quad 0.052 & \textbf{O4} &\quad -1.10 &\quad 0.157  &\textbf{O2} &\quad -0.833 &\quad 0.068\\[0.1cm]
\textbf{Ti1} &\quad 2.06 &\quad 0.051 & \textbf{O5} &\quad -0.90 &\quad 0.191  &\textbf{O3} &\quad -0.963 &\quad 0.132\\[0.1cm]
\textbf{Ti2} &\quad 2.06 &\quad 0.051 & \textbf{O6} &\quad -0.89 &\quad 0.201  &\textbf{O4} &\quad -0.949 &\quad 0.031\\[0.1cm]
 &  &  & \textbf{O7} &\quad -1.06 &\quad 0.046  &\textbf{O5} &\quad -1.05 &\quad 0.188\\[0.1cm]
 &  &  & \textbf{Ti1} &\quad 2.06 &\quad 0.046  &\textbf{O6} &\quad -1.07 &\quad 0.155\\[0.1cm]
 &  &  & & &  &\textbf{O7} &\quad -0.732&\quad -0.108\\[0.1cm]
 &  &  & & &  &\textbf{O8} &\quad -0.793&\quad -0.057\\[0.1cm]
 &  &  & & &  &\textbf{O9} &\quad -0.817&\quad 0.085\\[0.1cm]
 &  &  & & &  &\textbf{O10} &\quad -1.12&\quad -0.006\\[0.1cm]
 &  &  & & &  &\textbf{O11} &\quad -1.01&\quad 0.060\\[0.1cm]
 &  &  & & &  &\textbf{O12} &\quad -1.12&\quad 0.095\\[0.1cm]
 &  &  & & &  &\textbf{O13} &\quad -1.09&\quad 0.064\\[0.1cm]
 &  &  & & &  &\textbf{O14} &\quad -1.10&\quad 0.064\\[0.1cm]
 &  &  & & &  &\textbf{Ti1} &\quad 2.07&\quad 0.067\\[0.1cm]
&   &  & & &  &\textbf{Ti2} &\quad 2.06&\quad 0.045\\[0.1cm]
 &  &  & & &  &\textbf{Ti3} &\quad 2.05&\quad 0.047\\[0.1cm]
 &  &  & & &  &\textbf{Ti4} &\quad 2.06&\quad 0.044\\[0.3cm] \hline
\end{tabular}}
\vspace{0.8cm}
\caption*{\noindent Due to the charge redistribution the nearest atoms of Fe gained some amount of spin magnetization. The magnetic moment of O atoms of the cluster (O1-O3) were approximately 0.29-0.52\ $\mathrm{\mu_B}$ in the (\ch{Fe2O3})$_1$/\ch{TiO2}, 0.19-0.48\ $\mathrm{\mu_B}$ in the (\ch{Fe2O3})$_2$/\ch{TiO2} and -0.108-0.223\ $\mathrm{\mu_B}$ in the (\ch{Fe2O3})$_1$/\ch{TiO2}. At the surface, near the clusters, the spin magnetization of atoms was generally between 0.020 and 0.060 $\ \mathrm{\mu_B}$. The amount of magnetization spread around the rest of the atoms of \ch{TiO2} was negligible. The Bader charges of Ti atoms were around 2.0\ \textit{e}. This most probably indicates the oxidation state of Ti to be +4 \cite{55}. When forming bonds with the \ch{Fe2O3} clusters, Ti atoms tend to lose 2 electrons while the rest are involved in the covalent bonding. The charge redistribution of oxygen was generally around -1.0\  \textit{e} which can be identified as  oxidation state of -2 for oxygen \cite{55}.}
\label{Tab:heterostructures}
\end{table}




\vspace{1cm}
      
\begin{table}[h!]\centering
\caption{The bond lengths between the selected atoms in the (\ch{Fe2O3})$_1$/\ch{TiO2}-O$_\mathrm{v}$, (\ch{Fe2O3})$_1$/\ch{TiO2}-O$_\mathrm{sv3}$ and (\ch{Fe2O3})$_1$/\ch{TiO2}-O$_\mathrm{c}$. The labels of the atoms are shown in figure 8.}
\begin{tabular}{lll|ll|ll}
\hline \rule{0pt}{1.1\normalbaselineskip}& \multicolumn{2}{l|}{\textbf{(\ch{Fe2O3})$_1$/\ch{TiO2}-O$_\mathrm{v}$}} & \multicolumn{2}{l|}{\textbf{(\ch{Fe2O3})$_1$/\ch{TiO2}-O$_\mathrm{sv3}$}} & \multicolumn{2}{l}{\textbf{(\ch{Fe2O3})$_1$/\ch{TiO2}-O$_\mathrm{c}$}}\Tstrut \\[0.1cm]
\hline
 \rule{0pt}{1.5\normalbaselineskip}\makecell[l]{\textbf{Bond}\\\textbf{lengths (\AA)}}  & &    &   &    &  &   \\

\rule{0pt}{\normalbaselineskip}
&\quad \textbf{Fe1-O1} &\quad 1.85 &\quad \textbf{Fe1-O1} &\quad 2.06 &\quad \textbf{Fe1-O1} &\quad 1.89 \\[0.1cm]
&\quad \textbf{Fe1-O2} &\quad 1.85 &\quad \textbf{Fe1-O2} &\quad 1.91 &\quad \textbf{Fe1-O2} &\quad 1.89 \\[0.1cm]
&\quad \textbf{Fe1-O3} &\quad 1.81 &\quad \textbf{Fe1-O3} &\quad 1.85 &\quad \textbf{Fe1-O6} &\quad 1.86 \\[0.1cm]
&\quad \textbf{Fe2-O3} &\quad 1.86 &\quad \textbf{Fe2-O1} &\quad 2.21 &\quad \textbf{Fe2-O1} &\quad 2.06 \\[0.1cm]
&\quad \textbf{Fe2-O4} &\quad 1.95 &\quad \textbf{Fe2-O3} &\quad 1.80 &\quad \textbf{Fe2-O2} &\quad 2.06 \\[0.1cm]
&\quad \textbf{Fe2-O5} &\quad 1.95 &\quad \textbf{Fe2-O5} &\quad 1.95 &\quad \textbf{Fe2-O4} &\quad 1.96 \\[0.1cm]
&\quad \textbf{Ti1-O2} &\quad 1.85 &\quad \textbf{Ti1-O2} &\quad 1.79 &\quad \textbf{Fe2-O5} &\quad 1.96   \\[0.1cm]
&\quad \textbf{Ti2-O1} &\quad 1.85 &\quad \textbf{Ti2-O1} &\quad 1.83 &\quad \textbf{Ti1-O2} &\quad 1.94  \\[0.1cm]
& & & & &\quad  \textbf{Ti2-O1} &\quad 1.94 \\[0.3cm]  \hline        
\end{tabular}
\label{tab:O-vacs-bonds}
\end{table}

\vspace{0.5cm}

\begin{table}[h!]\centering
\caption{The Bader charges (in \textit{e}) and magnetic moments (in $\mathrm{\mu_B}$) of the (\ch{Fe2O3})$_1$/\ch{TiO2}-O$_\mathrm{v}$, (\ch{Fe2O3})$_1$/\ch{TiO2}-O$_\mathrm{sv3}$ and (\ch{Fe2O3})$_1$/\ch{TiO2}-O$_\mathrm{c}$. The labels of the atoms are shown in figure 8.}
\scalebox{1}{
\begin{tabular}{lll|lll|lll}
\hline    & \multicolumn{2}{l|}{\rule{0pt}{1.1\normalbaselineskip}\textbf{(\ch{Fe2O3})$_1$/\ch{TiO2}-O$_\mathrm{v}$}} &     & \multicolumn{2}{l|}{\textbf{(\ch{Fe2O3})$_1$/\ch{TiO2}-O$_\mathrm{sv3}$}} &     & \multicolumn{2}{l}{\textbf{(\ch{Fe2O3})$_1$/\ch{TiO2}-O$_\mathrm{c}$}} \\[0.2cm] \hline
 & \makecell[l]{\rule{0pt}{1.1\normalbaselineskip}\textbf{Bader}\\\textbf{charge}}   & \makecell[l]{\rule{0pt}{1.1\normalbaselineskip}\textbf{Magnetic}\\\textbf{moment}}  &     & \makecell[l]{\rule{0pt}{1.1\normalbaselineskip}\textbf{Bader}\\\textbf{charge}} & \makecell[l]{\rule{0pt}{1.1\normalbaselineskip}\textbf{Magnetic}\\\textbf{moment}} &     & \makecell[l]{\rule{0pt}{1.1\normalbaselineskip}\textbf{Bader}\\\textbf{charge}} & \makecell[l]{\rule{0pt}{1.1\normalbaselineskip}\textbf{Magnetic}\\\textbf{moment}}\Bstrut\\
\hline
\rule{0pt}{1.1\normalbaselineskip}\textbf{Fe1} & \hspace{0.2cm}1.52 & \hspace{0.4cm}4.02  & \textbf{Fe1} & \hspace{0.2cm}1.26 & \hspace{0.4cm}3.62              & \textbf{Fe1} & \hspace{0.2cm}1.29  & \hspace{0.4cm}3.67 \\[0.1cm]
\textbf{Fe2} & \hspace{0.2cm}1.30  & \hspace{0.4cm}3.77  & \textbf{Fe2} & \hspace{0.2cm}1.19 & \hspace{0.4cm}-3.57 & \textbf{Fe2} & \hspace{0.2cm}1.32 & \hspace{0.4cm}3.70 \\[0.1cm]
\textbf{O1}  & \hspace{0.2cm}-0.969  &\hspace{0.4cm} 0.104  & \textbf{O1} & \hspace{0.2cm}-1.00  & \hspace{0.4cm}-0.02 & \textbf{O1}  & \hspace{0.2cm}-1.07 & \hspace{0.4cm}0.072 \\[0.1cm]
\textbf{O2}  & \hspace{0.2cm}-0.967 & \hspace{0.4cm}0.104 & \textbf{O2} & \hspace{0.2cm}-0.985  & \hspace{0.4cm}-0.006 & \textbf{O2}  & \hspace{0.2cm}-1.08 & \hspace{0.4cm}0.072 \\[0.1cm]
\textbf{O3}  & \hspace{0.2cm}-1.01 & \hspace{0.4cm}0.373 & \textbf{O3}  & \hspace{0.2cm}-1.09 & \hspace{0.4cm}-0.003 & \textbf{O4}  & \hspace{0.2cm}-1.07 & \hspace{0.4cm}0.026  \\[0.1cm]
\textbf{O4} & \hspace{0.2cm}-1.08 & \hspace{0.4cm}0.032 & \textbf{O4}  & \hspace{0.2cm}-0.977  & \hspace{0.4cm}-0.002  & \textbf{O5}  & \hspace{0.2cm}-1.07  & \hspace{0.4cm}0.026  \\[0.1cm]
\textbf{O5}  & \hspace{0.2cm}-1.09  & \hspace{0.4cm}0.032 & \textbf{O5}  & \hspace{0.2cm}-1.10 & \hspace{0.4cm}-0.004 & \textbf{O6} & \hspace{0.2cm}-1.12 & \hspace{0.4cm}0.014 \\[0.1cm]
\textbf{O6}  & \hspace{0.2cm}-1.04 & \hspace{0.4cm} 0 & \textbf{O6} & \hspace{0.2cm}-1.05 & \hspace{0.4cm}0 & \textbf{Ti1} & \hspace{0.2cm}2.06 & \hspace{0.4cm}0.034 \\[0.1cm]
\textbf{Ti1} & \hspace{0.2cm}2.06 & \hspace{0.4cm}0.103 & \textbf{Ti1} & \hspace{0.2cm}2.01 & \hspace{0.4cm}0.072 & \textbf{Ti2} & \hspace{0.2cm}2.06 & \hspace{0.4cm}0.034 \\[0.1cm]
\textbf{Ti2} & \hspace{0.2cm}2.06 & \hspace{0.4cm}0.100 & \textbf{Ti2} & \hspace{0.2cm}2.04 & \hspace{0.4cm}0.033 &  &  & \\[0.3cm] \hline       
\end{tabular}}
\vspace{0.8cm}
\caption*{Due to the changes in the charge transfer the Bader charges of the Ti and O atoms nearest to Fe also changed. For oxygen the Bader charges further shows -2 oxidation state while +4 oxidation state for Ti. This also changed the magnetic moments. Generally, the spin magnetization of oxygen decreased in the presence of oxygen defect. In the presence of the O$_{\mathrm{v}}$ the Ti1 and Ti2 gained twice as much spin magnetization as in the defect-free heterostructure whereas in the (\ch{Fe2O3})$_1$/\ch{TiO2}-O$_\mathrm{c}$ their magnetic moments decreased. Due to the asymmetric structure of the (\ch{Fe2O3})$_1$ cluster Ti1 and Ti2 gained notable different amount of spin magnetization.}
\label{Charge-Mm-O-vacs}
\end{table}

\clearpage
